\documentclass[prd,twocolumn,floatfix,amsmath,nofootinbib,amssymb,floatfix]{revtex4}
\usepackage{graphicx,color,dcolumn,booktabs,bm}
\usepackage{longtable,lscape}
\usepackage{pdfpages}
\usepackage{txfonts}
\usepackage{overpic}
\usepackage{amssymb}
\usepackage{makecell}
\usepackage{indentfirst}
\usepackage{feynmf}
\usepackage{slashed}
\usepackage{cases}
\usepackage{color}
\usepackage{multirow}
\usepackage{threeparttable}
\usepackage{epstopdf}
\usepackage{enumerate}
\usepackage{subfigure}
\usepackage{textcomp}
\usepackage{diagbox}
\usepackage{mathrsfs}
\usepackage{cancel}
\usepackage{float}
\usepackage{amsmath}
\usepackage[misc]{ifsym}
\usepackage[colorlinks,
citecolor=blue,
anchorcolor=red,
menucolor=red,
linkcolor=red,
filecolor=red,
runcolor=red,
urlcolor=blue,
frenchlinks=red]{hyperref}
\usepackage{array}
\usepackage{booktabs}

\begin{document}

\title{The semileptonic decays of $\Omega_{b}^{*}$, $\Sigma_{b}^{*}$ and $\Xi_{b}^{\prime*}$ baryons}
	\author{Peng Yang$^{1}$}
	\author{Guo-Liang Yu$^{1,2}$}
	\email{yuguoliang2011@163.com}
	\author{Zhi-Gang Wang$^{1,2}$}
	\email{zgwang@aliyun.com}
	\author{Jie Lu$^{3}$}
	\author{Bin Wu$^{3}$}
	\affiliation{$^1$ Department of Mathematics and Physics, North China
		Electric Power University, Baoding 071003, People's Republic of
		China\\$^2$ Hebei Key Laboratory of Physics and Energy Technology, North China Electric Power University, Baoding 071000, China\\$^3$  School of Physics, Southeast University, Nanjing 210094, People’s Republic of China}
\date{\today}

\begin{abstract}

In this article, we employ the QCD sum rules basing on three-point correlation function to systematically analyze the transition form factors for the semileptonic decays of the spin \(\frac{3}{2}^{+}\) bottom baryons (\(\Omega_{b}^{*}\), \(\Sigma_{b}^{*}\), and \(\Xi_{b}^{\prime *}\)) to their corresponding spin \(\frac{1}{2}^{+}\) charmed partners. On the phenomenological side, we eliminate the contaminations of the baryons with negative parity and those with lower spin states. The operator product expansion is carried out up to dimension-six condensates, including the perturbative part, quark condensate \(\langle \bar{q} q\rangle\), gluon condensate \(\langle g_{s}^{2}GG\rangle\), mixed condensate \(\langle \bar{q} g_{s}\sigma Gq\rangle\), and four-quark condensates \(\langle \bar{q} q\rangle^{2}\) and \(g_{s}^{2}\langle \bar{q} q\rangle^{2}\). The form factors are evaluated in the space-like region and then are fitted into the time-like physical region via a $z$-series expansion approach. Finally, the semileptonic decay widths for $\Omega_{b}^{*}$, $\Sigma_{b}^{*}$ and $\Xi_{b}^{\prime*}$ baryons are calculated with our predicted form factors, and various polarization observables are also predicted. The predicted results in this work are compared with those predicted by other collaborations, and are expected to be useful for studying the properties of these singly heavy baryons.

\end{abstract}


\maketitle

\section{Introduction}\label{sec1}

In the last few decades, experimental and theoretical physicists have made great progresses in searching for the singly heavy baryons \cite{ParticleDataGroup:2026mpi}. Most of the ground states of singly heavy baryons were well established and some of the low-lying excited states were also detected. In the bottom baryon sector, the UA1 Collaboration firstly reported the observation of the $\Lambda_b$ baryon at the proton-antiproton collider at CERN \cite{UA1:1991vse}. In 2007, the D0 Collaboration observed another bottom baryon $\Xi_b$ in the $\Xi_b^- \to J/\psi \Xi^-$ \cite{D0:2007gjs} decay channel. After this observation, D0 Collaboration also detected the $\Omega_b^-$ baryon in the decay channel $\Omega_b^- \to J/\psi \Omega^-$ \cite{D0:2008sbw}. In 2012, an excited bottom baryon, denoted as $\Xi_b^{*0}$, was firstly observed by the CMS Collaboration according to the decay process $\Xi_b^{*0} \to \Xi_b^- \pi^+$ (plus charge conjugates) \cite{CMS:2012frl}. In addition, the spin doublet $\Sigma_b^{(\pm)}$ and $\Sigma_b^{*(\pm)}$ baryons were firstly detected by the CDF Collaboration in the invariant mass spectrum of $\Lambda_b^0 \pi^\pm$ \cite{CDF:2007oeq}. These experimental results have provided rich information for understanding the structure of bottom baryons. Up to now, the mass spectra of the bottom baryons were well studied by many collaborations \cite{Ebert:2011kk,Capstick:1986ter,Roberts:2007ni,Garcilazo:2007eh,Wang:2020mxk,Chen:2019ywy,Wang:2010fq,Wang:2009cr}. However, researches focused on their decay properties are limited, particularly concerning the weak decays.

The decay properties of bottom baryons constitute an important topic in hadron physics, and some theoretical works have been devoted to studying the strong, weak, and radiative decays \cite{Efimov:1991qj,Zhang:2026ugd,Niu:2026fzi,Luo:2025hjx,Wang:2025amv,Rivero-Acosta:2025drn,Wang:2024rai,Shu:2024jdn,Negash:2026orr,Patel:2026skc,Aliev:2009jt,Yao:2018jmc,Wang:2017kfr}. In recent years, weak decay property of heavy baryons has received particular attention, as precise experimental measurements can determine fundamental parameters of the weak decay process, such as CKM matrix elements. Reliable theoretical analysis is crucial to this problem, among which the accurate calculation of form factors is especially important. In Refs. \cite{Detmold:2015aaa,Meinel:2021rbm}, the form factors for transition processes $\Lambda_b \to p$ and $\Lambda_b \to \Lambda_c$ were predicted with the Lattice QCD method. In Ref. \cite{Lu:2026qkk}, the form factors of $\Sigma_b \to \Sigma_c$, $\Xi_b' \to \Xi_c'$, and $\Omega_b \to \Omega_c$ were also analyzed within the framework of three-point QCD sum rules. Besides, many other theoretical methods were also employed to carry out this work including the light-front quark model \cite{Ke:2019smy,Wang:2022ias}, light-cone QCD sum rules \cite{Aliev:2010uy,Azizi:2011mw}, the relativistic quark model \cite{Ebert:2006rp}, and the nonrelativistic quark model \cite{Cheng:1995fe}. It is noted that most of these works focus on the bottom baryons with the spin $J=\frac{1}{2}$ and the researches on $J=\frac{3}{2}$ states are very limited. Motivated by this situation, we will systematically investigate the semileptonic decays of spin $\frac{3}{2}$ bottom baryons $\Sigma_b^{*}$, $\Xi_b^{'*}$, and $\Omega_b^{*}$ to their spin $\frac{1}{2}$ charmed partners $\Sigma_c$, $\Xi_c'$, and $\Omega_c$. The QCD sum rules has been proven to be a powerful tool for studying hadronic properties such as masses, decay constants, and form factors \cite{Wang:2010it,Neishabouri:2025abl,Neishabouri:2024gbc,Luo:2025sns,Lu:2025gol,Yu:2026tbk,Wang:2026eci,Wang:2026dqi,Zhou:2025yjb}. Thus, this method will be employed to predict the form factors of these decay processes. We note that the semileptonic decays of $\Sigma_{b}^{*}$ and $\Omega_{b}^{*}$ have also been studied by Khajouei and Azizi within the method of three-point QCD sum rules \cite{Khajouei:2025tqw,Khajouei:2024frw}. In their studies, the value of each form factor is determined according to a selected Dirac structure, which may lead to the dependence of the results on structure choice. In the present work, we will employ all of the Dirac structures to construct linear equations to extract the form factors and to eliminate the structural dependence.

The paper is organized as follows. In Sec.~\ref{sec2}, we parametrize the hadronic matrix elements in terms of form factors and introduce how these form factors are extracted within the frame work of three-point QCD sum rules. In Sec.~\ref{sec3}, we present the numerical analysis, including the determination of the Borel window and prediction of the physical values of form factors by the $z$-series expansion. With these form factors, we then calculate the decay widths and various polarization observables for the semileptonic decay processes in Sec.~\ref{sec4}. We summarize our main conclusions in Sec.~\ref{sec5}. Some lengthy formulas are collected in Appendices \ref{Sec:AppA} and \ref{Sec:AppB}.

\section{QCD sum rules for the Transition Form Factors}\label{sec2}

The semileptonic decay of $\mathcal{B}_{b}^{*} \to \mathcal{B}_{c} l \bar{\nu}_l$ is related to the transition matrix element $\langle \mathcal{B}_{c}(p') | J_{\mu}^{V-A} | \mathcal{B}_{b}^{*}(p) \rangle$. This matrix element can be expressed in terms of four vector and axial-vector form factors as follows,
\begin{widetext}
\begin{eqnarray}\label{eq:1}
\notag
\left\langle \mathcal{B}_{c}\left(p^{\prime}\right)|J^{V-A}_{\mu}|\mathcal{B}_{b}^{*}\left(p\right) \right\rangle&&=\overline{u}^{\mathcal{B}_{c}}\left(p^{\prime},s^{\prime}\right)\left[g_{\mu\alpha}F_{1}\left(q^{2}\right)+\gamma_{\mu}\frac{p^{\prime}_{\alpha}}{m_{\mathcal{B}_{c}}}F_{2}\left(q^{2}\right)+\frac{p^{\prime}_{\alpha}p_{\mu}}{m_{\mathcal{B}_{c}}m_{\mathcal{B}_{b}^{*}}}F_{3}\left(q^{2}\right)+\frac{p^{\prime}_{\alpha}p^{\prime}_{\mu}}{m_{\mathcal{B}_{c}}^{2}}F_{4}\left(q^{2}\right)\right]\gamma_{5}u_{\alpha}^{\mathcal{B}_{b}^{*}}\left(p,s\right) \\
&&-\overline{u}^{\mathcal{B}_{c}}\left(p^{\prime},s^{\prime}\right)\left[g_{\mu\alpha}G_{1}\left(q^{2}\right)+\gamma_{\mu}\frac{p^{\prime}_{\alpha}}{m_{\mathcal{B}_{c}}}G_{2}\left(q^{2}\right)+\frac{p^{\prime}_{\alpha}p_{\mu}}{m_{\mathcal{B}_{c}}m_{\mathcal{B}_{b}^{*}}}G_{3}\left(q^{2}\right)+\frac{p^{\prime}_{\alpha}p^{\prime}_{\mu}}{m_{\mathcal{B}_{c}}^{2}}G_{4}\left(q^{2}\right)\right]u_{\alpha}^{\mathcal{B}_{b}^{*}}\left(p,s\right)
\end{eqnarray}
\end{widetext}
where $F_{i}$ and $G_{i}$ ($i=1\sim4$) are the vector and axial vector form factors, respectively. \(u(p^{\prime},s^{\prime})\) and \(u_{\alpha}(p,s)\) are the wave function which satisfy the Dirac and Rarita-Schwinger equations.
To calculate the form factors in Eq.~\eqref{eq:1}, we employ the three-point correlation function,
\begin{flalign}\label{eq:2}
&\Pi_{\mu\nu}(p,p')= & \notag \\
&i^2\int d^4x\,d^4y\,e^{ip'x}e^{i(p-p')y}\langle 0|T\{J^{\mathcal{B}_c}(x)J^{V-A}_{\mu}(y)\bar J^{\mathcal{B}_b^{*}}_{\nu}(0)\}|0\rangle &
\end{flalign}
On the phenomenological side, we insert a complete set of intermediate hadronic states with the same quantum numbers as the interpolating currents. This is achieved by utilizing the completeness relation \cite{Khodjamirian:2020btr,Colangelo:2000dp},
\begin{align}
1 =& |0\rangle\langle 0| + \sum_{h} \int \frac{d^4 k}{(2\pi)^4} 2\pi \delta(k^2 - m_h^2) |h(k)\rangle\langle h(k)| \notag \\
&+ \text{higher Fock states}
\label{eq:3}
\end{align}
where $|h(k)\rangle$ represents the intermediate hadronic state with momentum $k$.
Inserting Eq.~\eqref{eq:3} into Eq.~\eqref{eq:2} and using the double dispersion relation, we obtain the phenomenological side of the correlation function~\cite{Shifman:1978bx,Shifman:1978by},
\begin{align}
&\Pi_{\mu\nu}^{\mathrm{phy-V/A}}(p,p')
= \notag\\
&\frac{\langle 0 | J^{\mathcal{B}_c} | \mathcal{B}^{P_2}_c(p') \rangle \langle \mathcal{B}^{P_2}_c(p') | J^{V-A}_{\mu} | \mathcal{B}^{*P_1}_b(p) \rangle \langle \mathcal{B}^{*P_1}_b(p) | \bar{J}^{\mathcal{B}^{*}_b}_{\nu} | 0 \rangle}{(p'^2 - m_{\mathcal{B}^{P_2}_c}^2)(p^2 - m_{\mathcal{B}_b^{*P_1}}^2)}+ \notag\\
&\frac{\langle 0 | J^{\mathcal{B}_c} | \mathcal{B}^{P_2}_c(p') \rangle \langle \mathcal{B}^{P_2}_c(p') | J^{V-A}_{\mu} | \mathcal{B}^{P_1}_b(p) \rangle \langle \mathcal{B}^{P_1}_b(p) | \bar{J}^{\mathcal{B}^{*}_b}_{\nu} | 0 \rangle}{(p'^2 - m_{\mathcal{B}^{P_2}_c}^2)(p^2 - m_{\mathcal{B}_b^{P_1}}^2)} + \cdots
\label{eq:4}
\end{align}
where the ellipsis denotes the contributions from the excited and continuum states.
$\mathcal{B}_{b}^{P_{1}}$ and $\mathcal{B}_{b}^{*P_{1}}$ represent the mother particles with spins $J^{P_1}=\frac{1}{2}^{\pm}$ and $\frac{3}{2}^{\pm}$, respectively, while $\mathcal{B}_{c}^{P_2}$ is the daughter particle with spin $J^{P_2}=\frac{1}{2}^{\pm}$.
The hadronic transition elements $\langle \mathcal{B}^{P_2}_c(p') | J_{\mu}^{V-A} | \mathcal{B}^{P_1}_b(p)/\mathcal{B}^{*P_1}_b(p) \rangle$ in Eq.~\eqref{eq:4} can be parameterized in terms of a series of form factors.
The hadron vacuum matrix
elements for initial and final baryons are defined as,
\begin{eqnarray}\label{eq:5}
\notag
&&\left\langle \mathcal{B}_{b}^{*+}(p,s) \right|J^{\mathcal{B}_{b}^{*}}_{\nu}(0)\left| 0 \right\rangle  = \lambda_{\mathcal{B}_{b}^{*+}}u_{\nu}(p,s),\\
\notag
&&\left\langle \mathcal{B}_{b}^{*-}(p,s)  \right|J^{\mathcal{B}_{b}^{*}}_{\nu}(0)\left| 0\right\rangle  = \lambda_{\mathcal{B}_{b}^{*-}}u_{\nu}(p,s)i\gamma_{5},\\
\notag
&&\left\langle \mathcal{B}_{b}^{+}(p,s) \right|J_{\nu}^{\mathcal{B}_{b}^{*}}(0)\left| 0 \right\rangle  = \lambda_{\mathcal{B}_{b}^{+}}\Big(\alpha\gamma_{\nu}-\frac{4\alpha}{m}p_{\nu}\Big)u(p,s),\\
\notag
&&\left\langle \mathcal{B}_{b}^{-}(p,s) \right|J_{\nu}^{\mathcal{B}_{b}^{*}}(0)\left| 0 \right\rangle  = \lambda_{\mathcal{B}_{b}^{-}}\Big(\alpha\gamma_{\nu}-\frac{4\alpha}{m}p_{\nu}\Big)u(p,s)i\gamma_{5},\\
\notag
&&\left\langle 0\right|\overline{J}^{\mathcal{B}_{c}}\left|\mathcal{B}_{c}^{+}(p^{\prime},s^{\prime})\right\rangle=\lambda_{\mathcal{B}_{c}^{+}}\overline{u}(p^{\prime},s^{\prime}),\\
&&\left\langle 0\right|\overline{J}^{\mathcal{B}_{c}}\left|\mathcal{B}_{c}^{-}(p^{\prime},s^{\prime})\right\rangle=\lambda_{\mathcal{B}_{c}^{-}}\overline{u}(p^{\prime},s^{\prime})i\gamma_{5}.
\end{eqnarray}

It can be seen from Eq.~\eqref{eq:5} that the couplings of current \(J_{\nu}^{\mathcal{B}_{b}^{*}}\) to \(\mathcal{B}_{b}^{\pm}\) baryon with \(J^P = \frac{1}{2}^{\pm}\) are related to the structures \(\gamma_\nu\) and \(p_\nu\). After removing these structures and substituting the transition and vacuum matrix elements, the correlation function can be decomposed into the following different structures,
\begin{eqnarray}\label{eq:6}
\Pi_{\mu\nu}^{\mathrm{phy-V/A}}=\sum^{16}_{1}\Pi_{i}^{\mathrm{phy-V/A}}e_{i\mu\nu}^{\mathrm{V/A}},
\end{eqnarray}
where $\Pi_{i}^{\mathrm{phy-V/A}}$ represents the vector and axial-vector invariant amplitudes, and the form factors are contained in them. $e_{i\mu\nu}^{\mathrm{V/A}}$ denotes different dirac structures for the vector and axial vector parts. The vector part can be expressed as, $e_{i\mu\nu}^{\mathrm{V}}=\gamma_{5}g_{\mu\nu}, \slashed{p}\gamma_{5}g_{\mu\nu}, \slashed{p}^{\prime}\gamma_{5}g_{\mu\nu}, \slashed{p}\slashed{p}^{\prime}\gamma_{5}g_{\mu\nu}, \gamma_{\mu}\gamma_{5}p^{\prime}_{\nu}, \gamma_{\mu}\slashed{p}\gamma_{5}p^{\prime}_{\nu}, \\
 \gamma_{\mu}\slashed{p}^{\prime}\gamma_{5}p^{\prime}_{\nu}, \gamma_{\mu}\slashed{p}\slashed{p}^{\prime}\gamma_{5}p^{\prime}_{\nu}, \gamma_{5}p_{\mu}p^{\prime}_{\nu}, \slashed{p}\gamma_{5}p_{\mu}p^{\prime}_{\nu}, \slashed{p}^{\prime}\gamma_{5}p_{\mu}p^{\prime}_{\nu}, \slashed{p}\slashed{p}^{\prime}\gamma_{5}p_{\mu}p^{\prime}_{\nu}, \\
 \gamma_{5}p^{\prime}_{\mu}p^{\prime}_{\nu}, \slashed{p}\gamma_{5}p^{\prime}_{\mu}p^{\prime}_{\nu}, \slashed{p}^{\prime}\gamma_{5}p^{\prime}_{\mu}p^{\prime}_{\nu}, \slashed{p}\slashed{p}^{\prime}\gamma_{5}p^{\prime}_{\mu}p^{\prime}_{\nu}$.
The axial-vector structures $e_{i\mu\nu}^{A}$ can be obtained by removing the $\gamma_{5}$ from these above expressions.

To compute the QCD representation of the three-point correlation function in Eq.~\eqref{eq:2}, we first need to construct suitable interpolating currents for the heavy baryons involved. In the present work, we adopt the following interpolating currents,
\begin{eqnarray}\label{eq:7}
\notag
&&J^{\Sigma_{b}^{*0}}_{\nu}= \varepsilon _{abc}\frac{1}{\sqrt{2}}\left({u^{aT}\mathcal{C}\gamma _{\nu}d^b}+{d^{aT}\mathcal{C}\gamma _{\nu}u^b}\right)b^c, \\
\notag
&&J^{\Xi_{b}^{\prime*-}}_{\nu}= \varepsilon _{abc}\frac{1}{\sqrt{2}}\left({q^{aT}\mathcal{C}\gamma _{\nu}s^b}+{s^{aT}\mathcal{C}\gamma _{\nu}q^b}\right)b^c, \\
\notag
&&J^{\Omega_{b}^{*-}}_{\nu}= \varepsilon _{abc}\left({s^{aT}\mathcal{C}\gamma _{\nu}s^b}\right)b^c, \\
\notag
&&J^{\Sigma_{c}^{+}}= \varepsilon _{nml}\frac{1}{\sqrt{2}}\left({u^{nT}\mathcal{C}\gamma _{\alpha}d^m}+{d^{nT}\mathcal{C}\gamma _{\alpha}u^m}\right)\gamma_{\alpha}\gamma_{5}c^l, \\
\notag
&&J^{\Xi_{c}^{\prime0}}= \varepsilon _{nml}\frac{1}{\sqrt{2}}\left({q^{nT}\mathcal{C}\gamma _{\alpha}s^m}+{s^{nT}\mathcal{C}\gamma _{\alpha}q^m}\right)\gamma_{\alpha}\gamma_{5}c^l, \\
&&J^{\Omega_{c}^{0}}= \varepsilon _{nml}\left({s^{nT}\mathcal{C}\gamma _{\alpha}s^m}\right)\gamma_{\alpha}\gamma_{5}c^l
\end{eqnarray}
where $\varepsilon_{abc}$ is the Levi-Civita tensor, $\mathcal{C}$ is the charge conjugation matrix, and $q$ denotes a light quark field with $q=u$ or $d$.

Substituting the interpolating currents in Eq.~\eqref{eq:2} with their explicit forms in Eq.~\eqref{eq:7} and performing the Wick contractions, the correlation function on the QCD side is expressed as,
\begin{align}
&\Pi_{\mu\nu}^{\mathrm{QCD}}(p,p')
= 2i^{2}\varepsilon_{nml}\varepsilon_{abc}\int d^4x\,d^4y\, e^{ip'x}e^{iqy} \notag\\
&\quad \Big\{ \operatorname{Tr}\left[\mathcal{C}q^{naT}(x)\mathcal{C}\gamma_{\alpha}q^{\prime mb}(x)\gamma_{\nu}\right]\gamma_{\alpha}\gamma_{5}Q^{li}(x-y)\gamma_{\mu}Q^{i^{\prime}c}(y) \Big\}
\label{eq:8}
\end{align}
where \(Q^{ij}(x)\) and \(q^{ij}(x)\) are the heavy and light quark full propagators. By using the double dispersion relation, the correlation function on the QCD side can be written as,
\begin{eqnarray}\label{eq:9}
\Pi _{\mu\nu}^{\mathrm{QCD}}(p,p') &&= \int\limits_{s_{\min}}^\infty  {ds} \int\limits_{u_{\min}}^\infty  {du} \frac{{\rho _{\mu\nu} ^{\mathrm{QCD}}(s,u,q^2)}}{{(s - p^2)(u - p'^2)}}
\end{eqnarray}
where $\rho^{\mathrm{QCD}}_{\mu\nu}(s,u,q^2)$ is the QCD spectral density with $s=p^2$ and $u=p'^2$. For the creation thresholds $s_{\min}$ and $u_{\min}$ of the initial and final baryons, their values are taken to be the square of the sum of the masses of constituent quarks. The correlation functions on the QCD side can also be decomposed into the same structures as those on the phenomenological side,
\begin{eqnarray}\label{eq:10}
\Pi_{\mu\nu}^{\mathrm{QCD-V/A}}=\sum^{16}_{1}\Pi_{i}^{\mathrm{QCD-V/A}}e_{i\mu\nu}^{\mathrm{V/A}}
\end{eqnarray}
The scalar invariant amplitudes $\Pi_{\mu\nu}^{\mathrm{QCD-V/A}}$ on the QCD side can be represented as the summation of perturbative part and different vacuum condensate terms.
\begin{figure*}[htbp]
{\includegraphics[width=1\textwidth]{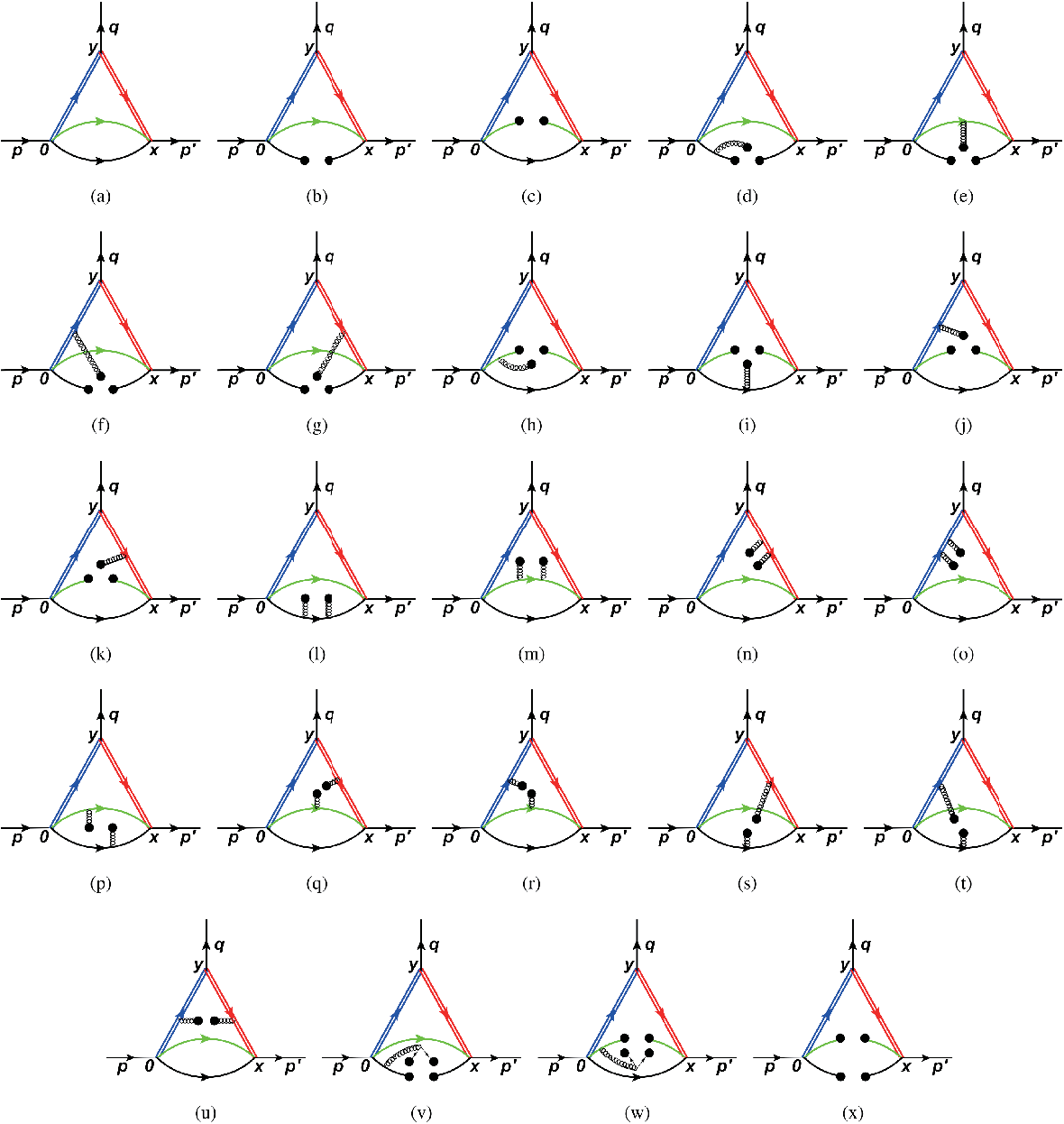}}
\caption{ The Feynman diagrams for the perturbative part and vacuum
condensate terms, where the doubly-solid line denotes a heavy quark, and the ordinary solid line represent a light quark.}
\label{feynman}
\end{figure*}
In the following, we take the perturbative term and the four quark condensate as examples to illustrate the calculation.

To compute the perturbative contribution, we first substitute the quark propagators with their free forms in momentum space. The corresponding Feynman diagram is shown in Fig.~\ref{feynman}(a). After performing the coordinate integrals, the perturbative part can be written as,
\begin{flalign}
\label{eq:11}
\notag
&\Pi_{\mu\nu}^{\mathrm{QCD-V0}}(p,p')= \frac{12}{{(2\pi)^8}}
\int d^4k_1\,d^4k_2\,d^4k_3\,d^4k_4 \int d^4q' &\\
&\quad \times
\frac{\delta^4(q'-k_1-k_2)\delta^4(p'-q'-k_3)\delta^4(q+k_3-k_4)}
{(k_1^2-m_1^2)(k_2^2-m_2^2)(k_3^2-m_3^2)(k_4^2-m_4^2)}
\, N_{\mu\nu} \;, &
\end{flalign}
where $N_{\mu\nu}=\operatorname{Tr}[(\slashed{k}_1-m_1)\gamma_\alpha(\slashed{k}_2+m_2)\gamma_\nu]\,
\gamma_\alpha\gamma_5(\slashed{k}_3+m_3)\gamma_\mu(\slashed{k}_4+m_4)$.
We apply the Cutkosky rule~\cite{Cutkosky:1960sp} by putting all quark lines on shell and the QCD spectral density for perturbative term can be expressed as,
\begin{flalign}
\label{eq:12}
\notag
&\rho_{\mu\nu}^{\mathrm{QCD-V0}}(s,u,q^2)
= \frac{12}{{(2\pi)^8}}\frac{(-2\pi i)^5}{(2\pi i)^3}
\int_{(m_1+m_2)^2}^{(\sqrt{u}-m_3)^2} dr' &\\ \notag
&\times
\int d^4k_3
\delta[(p'-k_3)^2-r']\,\delta(k_3^2-m_3^2)\,
\delta[(k_3+p-p')^2-m_4^2], &\\
&\times\int d^4k_1\delta(k_1^2-m_1^2)\delta[(q'-k_1)^2-m_2^2]N_{\mu\nu} &
\end{flalign}
with $q'=k_1+k_2$ and $r'=q'^2$. After finishing the integrations over the $\delta$ functions, the explicit form of the spectral density can be obtained. For the higher dimensional terms, namely the quark condensate (dimension 3), the gluon condensate (dimension 4), and the mixed condensate (dimension 5), we can also obtain their spectral densities according to similar calculation. The Feynman diagrams for these condensate terms are explicitly shown in Figs.~\ref{feynman}(b)$\sim$(c), Figs.~\ref{feynman}(l)$\sim$(u), and Figs.~\ref{feynman}(d)$\sim$(k), respectively.

The four quark condensate has two forms. The first form, shown in Fig.~\ref{feynman}(v)$\sim$(w), arises from a light quark line emitting a quark--antiquark pair together with a gluon that subsequently splits into another pair. This contribution is evaluated by replacing \(x_\mu\) appearing in the full propagator with \(i\,\partial/\partial p'_\mu\). Its explicit form is,
\begin{widetext}
\begin{equation}
\begin{aligned}
\label{eq:13}
\Pi_{\mu\nu 1}^{\mathrm{QCD-V6}}(p,p')
&= \frac{i g_s^2 \langle \bar q q \rangle^2}{648(2\pi)^4}
\Bigg\{
\frac{\partial}{\partial d}
\int d^4 k_3\,
\frac{N^{V6}_{\mu\nu 1}}
{\big[(p'-k_3)^2-d\big](k_3^2-m_3^2)\big[(p-p'+k_3)^2-m_4^2\big]} \\
&\quad + \frac{\partial^2}{2\partial d^2}
\int d^4 k_3\,
\frac{N^{V6}_{\mu\nu 2}}
{\big[(p'-k_3)^2-d\big](k_3^2-m_3^2)\big[(p-p'+k_3)^2-m_4^2\big]} \\
&\quad + \frac{\partial^3}{6\partial d^3}
\int d^4 k_3\,
\frac{N^{V6}_{\mu\nu 3}}
{\big[(p'-k_3)^2-d\big](k_3^2-m_3^2)\big[(p-p'+k_3)^2-m_4^2\big]}
\Bigg\}_{d\to m_1^2}
\end{aligned}
\end{equation}
\end{widetext}
with
\begin{flalign}
\label{eq:14}
&N^{V6}_{\mu\nu 1}
= 96\,g_{\alpha\nu}\,\gamma_{\beta}\gamma_5\,
(\slashed{k}_3+m_3)\gamma^{\mu}
(\slashed{p}-\slashed{p}'+\slashed{k}_3-m_4) \notag \\
&N^{V6}_{\mu\nu 2}
= -128\Bigl[
g^{\eta\nu}\Bigl( k_3\cdot p' (48m_1+63m_2)\notag \\
&
-8(3m_1+5m_2)\bigl(m_3^2+(p')^2\bigr)\Bigr)
+16(3m_1+4m_2)k_3^{\eta}(k_3^{\nu}-(p')^{\nu})\notag &\\
&
-16(3m_1+4m_2)(p')^{\eta}(k_3^{\nu}-(p')^{\nu})
\Bigr] \cdot \gamma^{\eta}\gamma_5\,(\slashed{k}_3+m_3)
\gamma^{\mu}\notag \\
&\times(\slashed{p}+\slashed{k}_3+m_4-\slashed{p}')\notag \\
&N^{V6}_{\mu\nu 3}
= 192(m_1+m_2)\bigl(2k_3\cdot p'-m_3^2-(p')^2\bigr)\notag \\
&\Bigl[
-g^{\eta\nu}\bigl(-2k_3\cdot p'+m_3^2+(p')^2\bigr)
+2k_3^{\eta}(k_3^{\nu}-(p')^{\nu}) \notag &\\
& -2(p')^{\eta}(k_3^{\nu}-(p')^{\nu})
\Bigr]\cdot \gamma^{\eta}\gamma_5\,(\slashed{k}_3+m_3)
\gamma^{\mu}(\slashed{p}+\slashed{k}_3+m_4-\slashed{p}')&
\end{flalign}
Again, we can also obtain its QCD spectral density according to a similar process as that of perturbative part.

The second form, shown in Fig.~\ref{feynman}(x), involves two independent light quark lines each producing a quark-antiquark pair. After performing the integration, this contribution can be expressed as,
\begin{flalign}
\label{eq:15}
\notag
\Pi_{\mu\nu 2}^{\mathrm{QCD-V6}}(p,p')
&= -\frac{\langle \bar q q \rangle^2}{12}
\frac{1}{p'^2-m_3^2}
\frac{1}{p^2-m_4^2}\\
&\times\operatorname{Tr}[\gamma_{\alpha}\gamma_{\nu}]\,
\gamma_{\alpha}\gamma_5
(\slashed{p}'+m_3)
\gamma_{\mu}
(\slashed{p}+m_4) \;, &
\end{flalign}
Its spectral density can easily be obtained by using the Cutkosky rule.

To improve the convergence of the quark-hadron duality and suppress the contributions from higher resonances and continuum, we perform double Borel transformations on both the phenomenological and QCD sides. The Borel transformation is defined as~\cite{Bracco:2011pg},
\begin{flalign}
\label{eq:16}
\mathcal{B}_{M^2}\left[\frac{1}{(m^2-q^2)^n}\right]
= \frac{1}{(n-1)!}\frac{e^{-m^2/M^2}}{(M^2)^{n-1}}, \qquad n>0 \; &
\end{flalign}
Then, performing the change of variables $p^2 \to -P^2$, $p'^2 \to -P'^2$, $q^2 \to -Q^2$ on both sides and applying the double Borel transformations with respect to $P^2$ and $P'^2$, the square of momentum will be replaced by the Borel parameters $M_1^2$ and $M_2^2$. Since there are 16 Dirac structures on both the phenomenological and QCD sides, the quark hadron duality yields 16 linear equations about the form factors. By solving these equations, we obtain the momentum dependent form factors $F_i(Q^2)$ and $G_i(Q^2)$, whose explicit expressions are collected in Appendix~\ref{Sec:AppA}.

\section{Numerical results and Discussions}\label{sec3}
To obtain the numerical results, the values of some input parameters such as the baryon masses, pole residues, quark masses, and various vacuum condensates, need to be determined.
All of the parameters used in the present work are listed in Table~\ref{PV}.\begin{table*}[htbp]
\caption{Input parameters used in the present work.}
\begin{ruledtabular}
\renewcommand{\arraystretch}{1.3}
\label{PV}
\begin{tabular}{c| c c c c c c}
IP&$m_{\Sigma_{c}(\frac{1}{2}^{+})}$ \cite{Yu:2022ymb}&$m_{\Sigma_{c}(\frac{1}{2}^{-})}$ \cite{Yu:2022ymb}& $m_{\Omega_{c}(\frac{1}{2}^{+})}$ \cite{Yu:2022ymb}&$m_{\Omega_{c}(\frac{1}{2}^{-})}$  \cite{Yu:2022ymb}& $m_{\Xi_{c}^{\prime}(\frac{1}{2}^{+})}$ \cite{Li:2024zze}& $m_{\Xi_{c}^{\prime}(\frac{1}{2}^{-})}$ \cite{Li:2024zze}\\
Values(GeV)& $2.457$ &$2.823$ &$2.699$ &$3.057$ & $2.59$& $2.927$\\ \hline
IP&$m_{\Sigma_{b}^{*}(\frac{3}{2}^{+})}$ \cite{Yu:2022ymb}&$m_{\Sigma_{b}^{*}(\frac{3}{2}^{-})}$ \cite{Yu:2022ymb}&$m_{\Omega_{b}^{*}(\frac{3}{2}^{+})}$ \cite{Yu:2022ymb}& $m_{\Omega_{b}^{*}(\frac{3}{2}^{-})}$ \cite{Yu:2022ymb}&$m_{\Xi_{b}^{*\prime}(\frac{3}{2}^{+})}$  \cite{Li:2024zze}& $m_{\Xi_{b}^{*\prime}(\frac{3}{2}^{-})}$  \cite{Li:2024zze}\\
Values(GeV)& $5.849$ &$6.116$ & $6.069$& $6.336$&$5.971$&$6.233$\\ \hline
IP&$\lambda_{\Sigma_{c}(\frac{1}{2}^{+})}$ \cite{Wang:2009cr}&$\lambda_{\Omega_{c}(\frac{1}{2}^{+})}$ \cite{Wang:2009cr}&$\lambda_{\Xi_{c}^{\prime}(\frac{1}{2}^{+})}$  \cite{Wang:2009cr}& $\lambda_{\Sigma_{b}^{*}(\frac{3}{2}^{+})}$ \cite{Wang:2010vn}& $\lambda_{\Omega_{b}^{*}(\frac{3}{2}^{+})}$ \cite{Wang:2010vn}& $\lambda_{\Xi_{b}^{*\prime}(\frac{3}{2}^{+})}$ \cite{Wang:2010vn}\\
Values(GeV$^{3}$)& $\sqrt{2}\times(0.045\pm0.015)$ &$0.093\pm0.023$ &$\sqrt{2}\times(0.055\pm0.016)$ &$\sqrt{2}\times(0.038\pm0.011)$ &$0.083\pm0.018$&$\sqrt{2}\times(0.049\pm0.012)$ \\ \hline
IP& $m_{e}$ \cite{ParticleDataGroup:2024cfk} & $m_{\mu}$ \cite{ParticleDataGroup:2024cfk}& $m_{\tau}$ \cite{ParticleDataGroup:2024cfk} & $\langle \overline{q}q\rangle$ \cite{Lee:2022jjn,Reinders:1984sr} & $\langle \overline{s}s\rangle$  \cite{Lee:2022jjn,Reinders:1984sr} & $\langle \overline{q}g_{s}\sigma Gq\rangle$ \cite{Lee:2022jjn,Reinders:1984sr} \\
Values& $0.511\times10^{-3}$ GeV&$106\times10^{-3}$ GeV&1.78 GeV& $-(0.23\pm0.01)^{3}$ GeV$^{3}$& $(0.8\pm0.1)\langle \overline{q}q\rangle$&$m_{0}^{2}\langle \overline{q}q\rangle$\\ \hline
IP&$\langle \overline{s}g_{s}\sigma Gs\rangle$ \cite{Lee:2022jjn,Reinders:1984sr} & $m_{0}^{2}$ \cite{Lee:2022jjn,Reinders:1984sr} & $\langle g_{s}^{2}GG\rangle$  \cite{Narison:2010cg,Narison:2011xe,Narison:2011rn} && &\\
Values&$m_{0}^{2}\langle \overline{s}s\rangle$& $0.8\pm0.1$ GeV$^{2}$&$0.47\pm0.15$ GeV$^{4}$&& &\\
\end{tabular}
\end{ruledtabular}
\end{table*} It is known that the values of heavy quark masses and vacuum condensates depend on the energy scale. According to the renormalization group equation (RGE), these dependencies are,
\begin{eqnarray}\label{eq:17}
\notag
&&m_{c[b]}(\mu ) = m_{c[b]}(m_{c[b]})\left[\frac{\alpha _s(\mu )}{\alpha _s(m_{c[b]})}\right]^{\frac{12}{33 - 2n_f}}\\
\notag
&&m_{s}(\mu ) = m_{s}(2\mathrm{GeV})\left[\frac{\alpha _s(\mu )}{\alpha _s(2\mathrm{GeV})}\right]^{\frac{12}{33 - 2n_f}}\\
\notag
&&\left\langle \bar qq \right\rangle (\mu ) = \left\langle \bar qq \right\rangle (1\mathrm{GeV})\left[\frac{\alpha _s(1\mathrm{GeV})}{\alpha _s(\mu )}\right]^{\frac{12}{33 - 2n_f}}\\
\notag
&&\left\langle \bar ss \right\rangle (\mu ) = \left\langle \bar ss \right\rangle (1\mathrm{GeV})\left[\frac{\alpha _s(1\mathrm{GeV})}{\alpha _s(\mu )}\right]^{\frac{12}{33 - 2n_f}}\\
\notag
&&\left\langle \bar qg_s\sigma Gq \right\rangle (\mu ) = \left\langle \bar qg_s\sigma Gq \right\rangle (1\mathrm{GeV})\left[\frac{\alpha _s(1\mathrm{GeV})}{\alpha _s(\mu )}\right]^{\frac{2}{33 - 2n_f}}\\
\notag
&&\left\langle \bar sg_s\sigma Gs \right\rangle (\mu ) = \left\langle \bar sg_s\sigma Gs \right\rangle (1\mathrm{GeV})\left[\frac{\alpha _s(1\mathrm{GeV})}{\alpha _s(\mu )}\right]^{\frac{2}{33 - 2n_f}}
\end{eqnarray}
\begin{eqnarray}
\notag
\alpha _s(\mu ) = \frac{1}{b_0t}\left[ 1 - \frac{b_1}{b_0^2}\frac{\log t}{t}+ \frac{b_1^2(\log ^2t - \log t - 1) + b_0b_2}{b_0^4t^2} \right]\\
\end{eqnarray}
where $t=\log\frac{\mu^2}{\Lambda_{QCD}^2}$, $b_0=\frac{33-2n_f}{12\pi}$, $b_1=\frac{153-19n_f}{24\pi^2}$, $b_2=\frac{2857-\frac{5033}{9}n_f+\frac{325}{27}n_f^2}{128\pi^3}$, $\Lambda_{QCD}=213$ MeV for the quark flavors $n_f=5$  \cite{ParticleDataGroup:2024cfk}. The minimum subtraction masses of $c, b$ and $s$ quarks are taken from the Particle Data Group,
which are $m_c(m_c)=1.275\pm0.025$ GeV and $m_b(m_b)=4.18\pm0.03$ GeV and $m_s(\mu$ = 2GeV) = 0.095 $\pm$ 0.005 GeV. The energy
scale $\mu$ = 2 GeV is adopted for $\Sigma_b^{*}$,$\Omega_b^{*}$ and $\Xi_b^{\prime*}$
transitions in the present work.

\begin{figure}[htbp] \centering
{\includegraphics[width=0.5\textwidth]{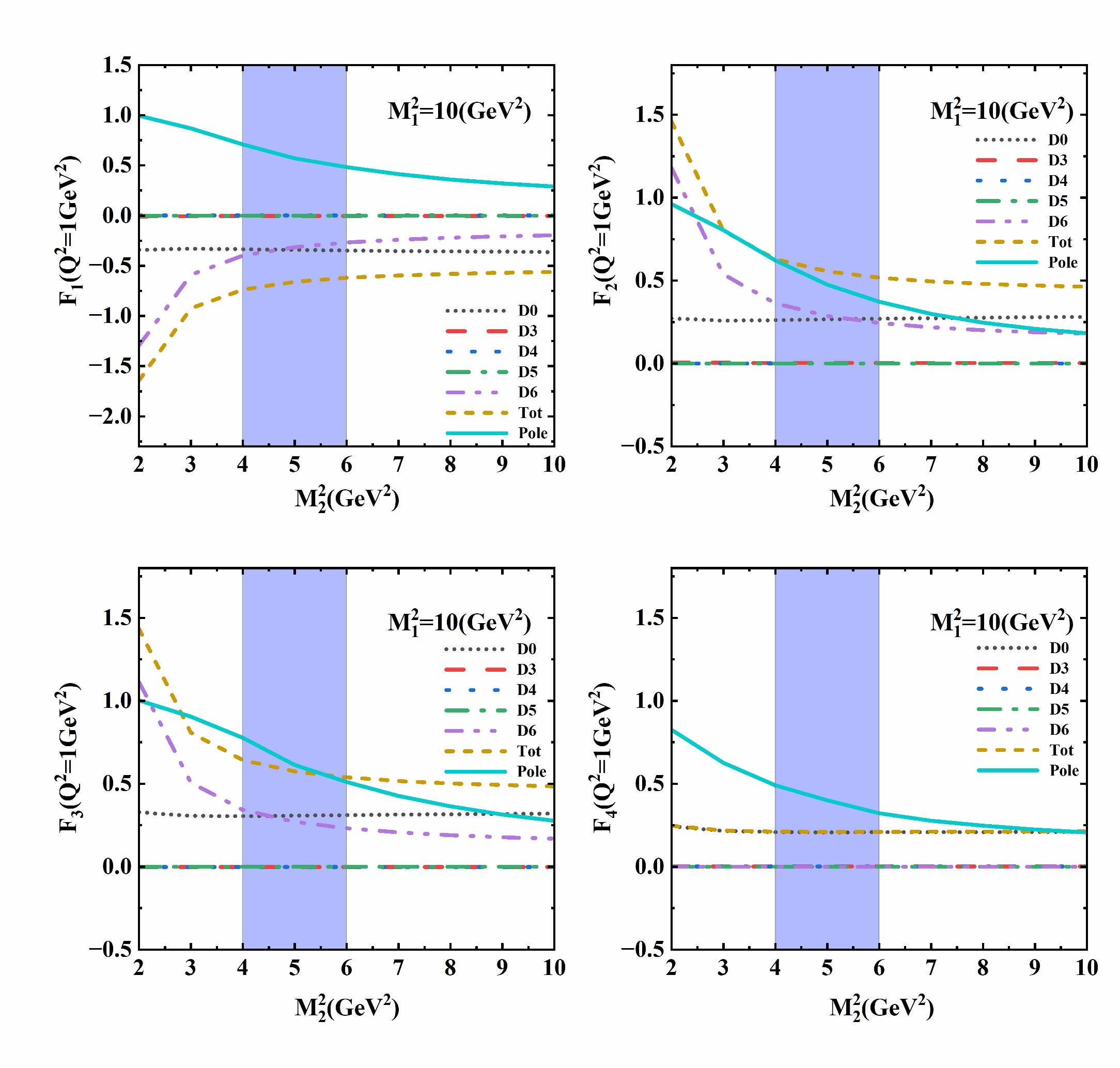}}
\caption{The pole contribution, perturbative term, and contributions of different vacuum condensates as functions of the Borel parameter $\mathrm{M}_{2}^{2}$ for the form factors $F_1$, $F_2$, $F_3$, and $F_4$ of the transition $\Sigma_b^{*0} \to \Sigma_c^{+}$. The grey areas indicate the Borel window.}
\label{pole}
\end{figure}

\begin{figure}[htbp] \centering{\includegraphics[width=0.5\textwidth]{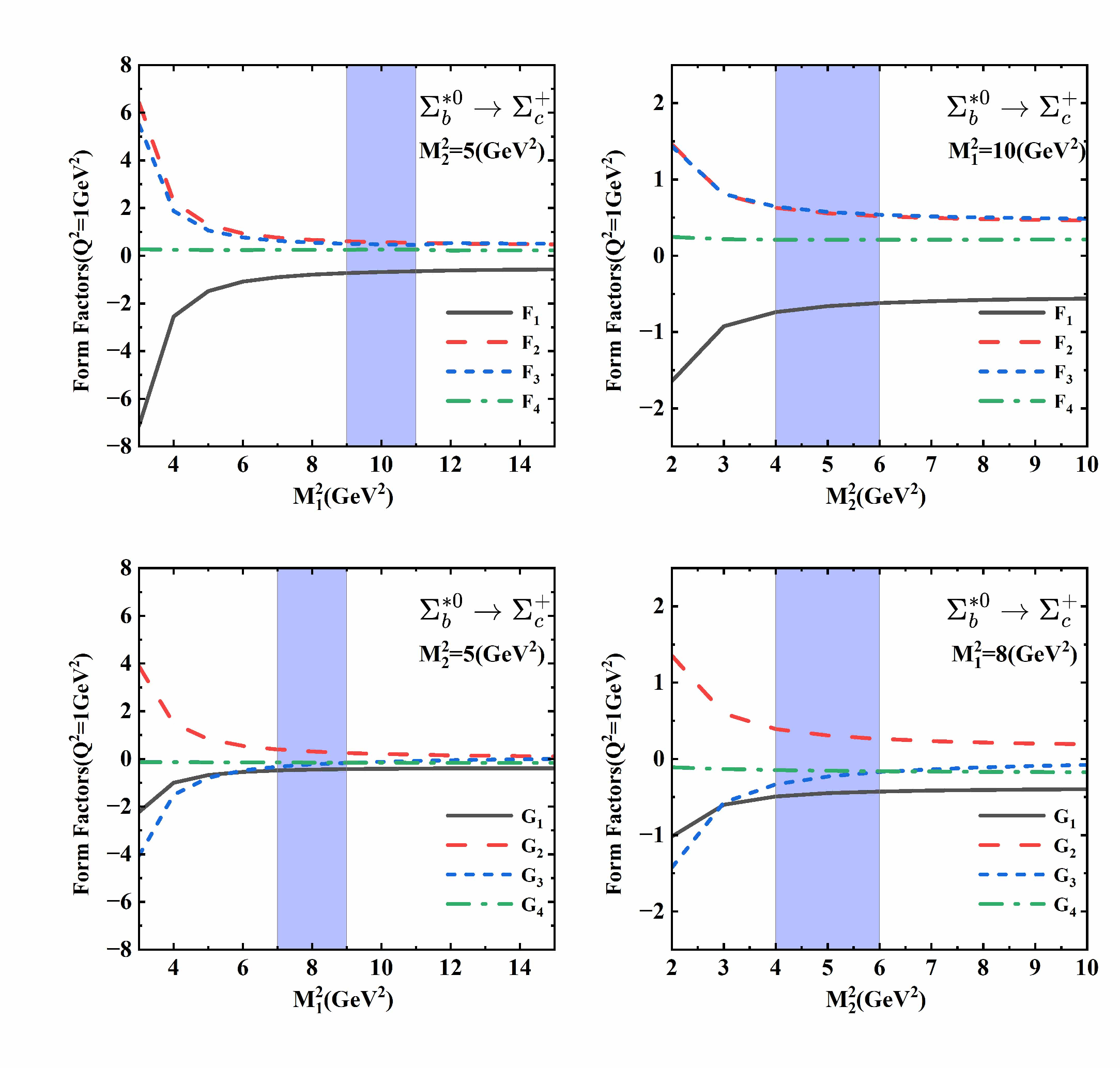}}
\caption{Variations of the form factors for the $\Sigma_{b}^{*0}\to\Sigma_{c}^{+}$ transition process with respect to the Borel parameters $\mathrm{M}_{1}^{2}$ and $\mathrm{M}_{2}^{2}$.}
\label{pole3}
\end{figure}

To obtain reliable results, an appropriate working region, named as Borel window, for the Borel parameters should be selected. In this region, the values of the form factors should have a weak dependence on the Borel parameters. Meanwhile, two criteria should be satisfied: pole dominance and convergence of the OPE. The pole contribution is defined as,
\begin{eqnarray}
\label{eq:18}
\mathrm{Pole}=\frac{\int_{s_{min}}^{s_{0}} ds\int_{u_{min}}^{u_{0}} du\rho^{\mathrm{QCD}}\left(s,u,q^{2}\right)\mathrm{exp}\left(-\frac{s}{\mathrm{M}_{1}^{2}}-\frac{u}{\mathrm{M}_{2}^{2}}\right)}{\int_{s_{min}}^{\infty} ds\int_{u_{min}}^{\infty} du\rho^{\mathrm{QCD}}\left(s,u,q^{2}\right)\mathrm{exp}\left(-\frac{s}{\mathrm{M}_{1}^{2}}-\frac{u}{\mathrm{M}_{2}^{2}}\right)}
\end{eqnarray}
For the vector form factors of transition process $\Sigma_{b}^{*0}\to\Sigma_{c}^{+}$ as an example, we introduce how the Borel window is determined. In Fig.~\ref{pole}, we show the pole contributions and the contributions of the perturbative part and various condensates as functions of the Borel parameter $M_{2}^{2}$. The grey area ($4\le M_2^2\le6$ GeV$^2$) indicates the selected Borel window. From this figure, we can see that both of the pole contribution and the contributions of high dimension condensates decrease with the Borel parameters. The criteria of pole dominance and convergence of the OPE require the pole contribution should be larger than $40\%$, at the same time, the contributions of high dimension condensates should be small or stable. It is shown from this figure that these two conditions are all well satisfied in the Borel window. Besides, we can also see that the contribution of four quark condensate (D=6) is significant and can not be omitted in the present work. In addition, we also plot the variations of the form factors $F_{i}/G_{i}$ ($i=1\sim4$) with respect to the Borel parameters \(M_1^2\) and \(M_2^2\) in Fig.~\ref{pole3}.
From this figure, we can see that the results show well stability and weak dependence on the Borel
parameters in the Borel platform.

\begin{figure}[htbp] \centering
{\includegraphics[width=0.5\textwidth]{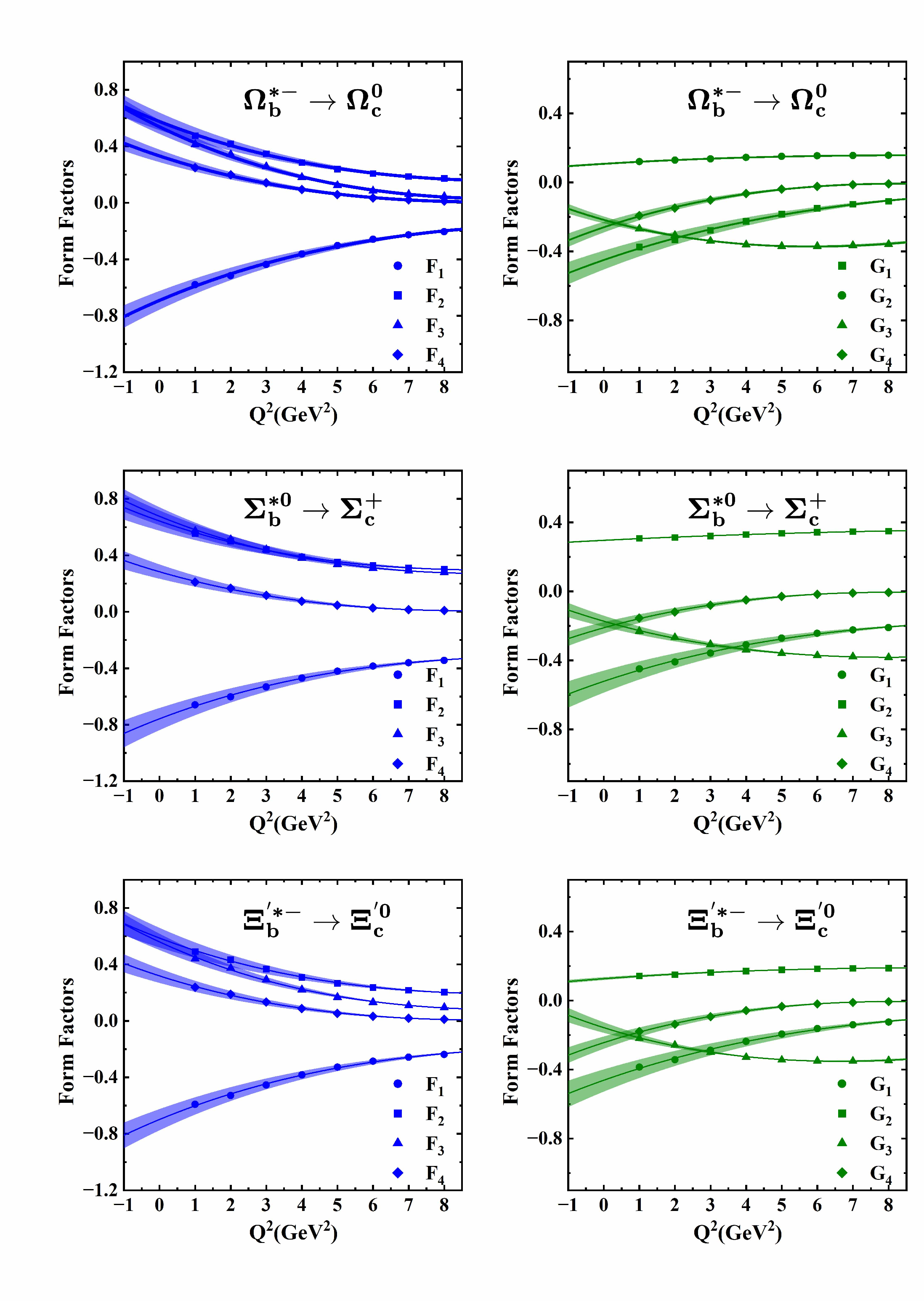}}
\caption{Numerical results and fitted graphs for transition processes $\Omega_b^{*-}\to\Omega_c^{0}$, $\Sigma_b^{*0}\to\Sigma_c^{+}$, and $\Xi_b^{\prime *-}\to\Xi_c^{'0}$ as functions of $Q^2$.}
\label{fitting1}
\end{figure}

After all of the conditions are satisfied, we can obtain the form factors in the space-like region by setting $Q^2=1\sim8$ GeV$^{2}$. To obtain the results in physical region $Q^2\leq0$, the numerical results should be fitted into analytical function and be extrapolated into the time-like region. In the present work, the $z$-series expansion is employed to realize this goal, and it can be expressed as,
\begin{eqnarray}
\notag
&&F_{i}(Q^2) = \\ \notag
&&\frac{F(0)}{1 + Q^2/m_{\mathrm{B_{c}^{*}}\left(1^{-}\right)}^{2}}\times\Bigg\{1+a\left[z(Q^{2})-z(0)-\frac{1}{3}[z(Q^2)^{3}-z(0)^3]\right]\\ \notag
&&+b\left[z(Q^{2})^{2}-z(0)^{2}-\frac{2}{3}[z(Q^2)^3-z(0)^3]\right]\Bigg\},
\end{eqnarray}
\begin{eqnarray}\label{eq:19}
\notag
&&G_{i}(Q^2) = \\ \notag
&&\frac{G(0)}{1 + Q^2/m_{\mathrm{B_{c}^{*}}\left(1^{+}\right)}^{2}}\times\Bigg\{1+c\left[z(Q^{2})-z(0)-\frac{1}{3}[z(Q^2)^{3}-z(0)^3]\right]\\
&&+d\left[z(Q^{2})^{2}-z(0)^{2}-\frac{2}{3}[z(Q^2)^3-z(0)^3]\right]\Bigg\},
\end{eqnarray}
with $z(Q^2)=\frac{\sqrt{\vphantom{Q^2 t_-} t_{+}+Q^{2}} - \sqrt{\vphantom{Q^2 t_-} t_{+}-t_{-}}
}{\sqrt{\vphantom{Q^2 t_-} t_{+}+Q^{2}} + \sqrt{\vphantom{Q^2 t_-} t_{+}-t_{-}}
}$ and $t_\pm=(m_{\mathcal B_b^*}\pm m_{\mathcal B_c})^2$.The masses of the vector and axial-vector $B_c$ mesons are taken from the modified Godfrey-Isgur quark model~\cite{Li:2023wgq}
$m_{B_c^*(1^-)} = 6.338\ \text{GeV},m_{B_c^*(1^+)} = 6.745\ \text{GeV}$.
The fitting results are explicitly shown in Fig.~\ref{fitting1}, where the shaded area denotes the uncertainties of the results. It is shown by these figures that the form factors are fitted well by the $z$-series expansion approach. The fitting parameters and the values of form factors at $Q^2=0$ are all listed in Table~\ref{FP}.
\begin{table*}[htbp]
\caption{Form factors at $Q^2=0$ and fitting parameters $a,b,c,d$ for the $z$-series expansion. It is noted that the values of the form factors for \(\Sigma_b^{*+}\to\Sigma_c^{++}\) and \(\Sigma_b^{*-}\to\Sigma_c^{0}\) are the same as those of \(\Sigma_b^{*0}\to\Sigma_c^{+}\). The values for $\Xi_{b}^{\prime*0}\rightarrow\Xi_{c}^{\prime+}$ is the same as those of $\Xi_{b}^{\prime*-}\rightarrow\Xi_{c}^{\prime0}$.}
\begin{ruledtabular}
\renewcommand{\arraystretch}{1.3}
\label{FP}
\begin{tabular}{c c c c c c c c c c c}
Mode&FF&$F_{i}(0)$&$a$&$b$&$F_{i}(0)$ \cite{Khajouei:2025tqw,Khajouei:2024frw}&FF&$G_{i}(0)$&$c$&$d$&$G_{i}(0)$ \cite{Khajouei:2025tqw,Khajouei:2024frw}\\
\hline
\multirow{4}{*} {$\Omega_{b}^{*-}\rightarrow\Omega_{c}^{0}$}&$F_{1}$&$0 .64_ {-0.052}^{+1.40}$ & $-105.18_ {-23.04}^{+33.02}$ & $0 .70_ {-0.21}^{+0.21}\times 10^{3}$&$-0.085_ {-0.002}^{+0.002}$ &
$G_{1}$& $0 .10_ {-0.30}^{+0.61}$ & $84 .97_ {-165.39}^{+212.41}$ & $-0.49_ {-0.95}^{+1.77}\times 10^{3}$&$-0 .059_ {-0.001}^{+0.001}$\\
&$F_{2}$&$-0.69_ {-1.07}^{+1.27}$ & $-82.19_ {-21.21}^{+78.19}$ & $0 .50_ {-0.19}^{+0.60}\times 10^{3}$&$-1.26_ {-0.25}^{+0.25}$ &
$G_{2}$&$-0.45_ {-0.16}^{+0.56}$ & $-81.31_ {-101.48}^{+168.95}$ & $0 .47_ {-0.83}^{+1.00}\times 10^{3}$&$0 .39_ {-0.08}^{+0.08}$\\
&$F_{3}$&$0 .33_ {-0.21}^{+0.96}$ & $-163.38_ {-22.82}^{+83.25}$ & $1 .13_ {-0.19}^{+0.63}\times 10^{3}$&$-0 .39_ {-0.08}^{+0.08}$ &
$G_{3}$&$-0.26_ {-0.044}^{+0.14}$ & $-186.50_ {-105.72}^{+433.60}$ & $1 .33_ {-0.86}^{+3.23}\times 10^{3}$&$-0 .41_ {-0.08}^{+0.08}$\\
&$F_{4}$&$0 .49_ {-0.027}^{+0.20}$ & $-141.27_ {-24.55}^{+41.46}$ & $0 .95_ {-0.20}^{+0.27}\times 10^{3}$&$-0 .31_ {-0.06}^{+0.06}$ &
$G_{4}$&$-0.24_ {-0.008}^{+0.35}$ & $201 .68_ {-119.10}^{+391.48}$ & $-1.56_ {-1.05}^{+2.92}\times 10^{3}$&$-0 .31_ {-0.06}^{+0.06}$\\
\hline
\multirow{4}{*} {$\Sigma_{b}^{*0}\rightarrow\Sigma_{c}^{+}$}&$F_{1}$&$0 .72_ {-0.02}^{+1.56}$ & $-84.46_ {-5.80}^{+16.47}$ & $526 .35_ {-29.50}^{+134.05}$&$-5.07_ {-0.93}^{+0.93}$ &
$G_{1}$&$0 .29_ {-0.43}^{+0.88}$ & $19 .02_ {-86.35}^{+306.65}$ & $-0.03_ {-0.40}^{+1.91}\times 10^{3}$&$8.17_ {-1.05}^{+1.05}$\\
&$F_{2}$&$-0.76_ {-1.09}^{+1.40}$ & $-64.33_ {-14.55}^{+90.44}$ & $376 .00_ {-121.49}^{+570.22}$&$-1.73_ {-0.40}^{+0.40}$ &
$G_{2}$&$-0.52_ {-0.28}^{+0.82}$ & $-65.42_ {-85.53}^{+107.65}$ & $0 .36_ {-0.41}^{+0.73}\times 10^{3}$&$-0 .48_ {-0.11}^{+0.11}$\\
&$F_{3}$&$0 .28_ {-0.39}^{+0.97}$ & $-157.03_ {-69.86}^{+96.98}$ & $963 .76_ {-421.07}^{+606.85}$&$-0 .95_ {-0.14}^{+0.14}$ &
$G_{3}$&$-0.21_ {-0.04}^{+0.25}$ & $-176.65_ {-114.44}^{+411.83}$ & $1 .12_ {-0.77}^{+2.52}\times 10^{3}$&$-0 .49_ {-0.09}^{+0.09}$\\
&$F_{4}$&$0 .61_ {-0.04}^{+0.38}$ & $-82.99_ {-10.82}^{+76.64}$ & $523 .33_ {-61.26}^{+460.86}$&$-0 .44_ {-0.09}^{+0.09}$ &$G_{4}$&$-0.21_ {-0.03}^{+0.51}$ & $170 .77_ {-149.38}^{+350.81}$ & $-1.01_ {-0.95}^{+2.15}\times 10^{3}$&$-0 .44_ {-0.09}^{+0.09}$\\
\hline \multirow{4}{*}{$\Xi_{b}^{\prime*-}\rightarrow\Xi_{c}^{\prime0}$}&$F_{1}$&$0 .66_ {-0.04}^{+1.43}$ & $-98.82_ {-19.77}^{+24.95}$ & $0 .64_ {-0.15}^{+0.17}\times 10^{3}$&-&
$G_{1}$&$0 .12_ {-0.24}^{+0.65}$ & $91 .06_ {-170.70}^{+382.46}$ & $-0.49_ {-0.93}^{+2.80}\times 10^{3}$&-\\
&$F_{2}$&$-0.70_ {-1.07}^{+1.29}$ & $-77.03_ {-18.21}^{+81.70}$ & $0 .46_ {-0.16}^{+0.58}\times 10^{3}$&-&
$G_{2}$&$-0.46_ {-0.18}^{+0.59}$ & $-79.60_ {-101.82}^{+161.62}$ & $0 .45_ {-0.78}^{+0.90}\times 10^{3}$&-\\
&$F_{3}$&$0 .32_ {-0.24}^{+0.94}$ & $-161.38_ {-37.42}^{+87.43}$ & $1 .06_ {-0.27}^{+0.61}\times 10^{3}$&-&
$G_{3}$&$-0.24_ {-0.09}^{+0.16}$ & $-184.93_ {-107.07}^{+520.57}$ & $1 .26_ {-0.81}^{+3.60}\times 10^{3}$&-\\
&$F_{4}$&$0 .50_ {-0.02}^{+0.23}$ & $-122.21_ {-32.29}^{+41.51}$ & $0 .79_ {-0.20}^{+0.29}\times 10^{3}$&-&
$G_{4}$&$-0.19_ {-0.02}^{+0.32}$ & $245 .44_ {-171.60}^{+433.39}$ & $-1.72_ {-1.30}^{+3.00}\times 10^{3}$ &-
\end{tabular}
\end{ruledtabular}
\end{table*}

\begin{figure}[htbp] \centering
{\includegraphics[width=0.5\textwidth]{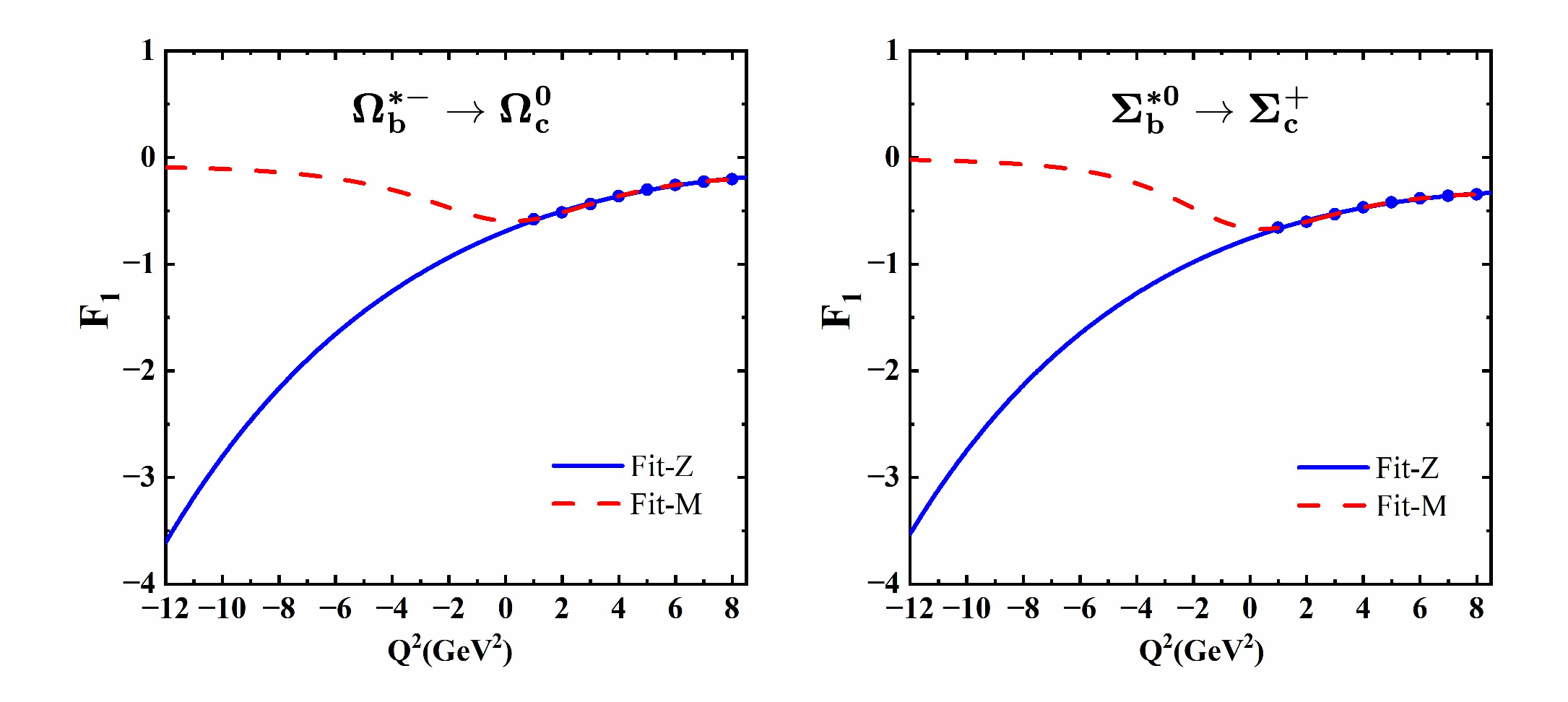}}
\caption{Comparison between the z series expand approach (solid blue line) and the multipole method (dashed red line).}
\label{zazz}
\end{figure}

From Table~\ref{FP}, we observe that our results differ greatly from those in Refs.~\cite{Khajouei:2025tqw} and~\cite{Khajouei:2024frw}. This difference may be related to the following two reasons. Firstly, it is known from Eq. (\ref{eq:4}) that the current \(J_{\nu}^{\mathcal{B}_{b}^{*}}\) couples not only to the $J^{P}=\frac{3}{2}^{\pm}$ states but also to $\frac{1}{2}^{\pm}$ ones, and the current $J^{\mathcal{B}_c}$ couples to both of the $\frac{1}{2}^{+}$ and $\frac{1}{2}^{-}$ states. We considered all of these couplings and successfully eliminate contaminations of the $\frac{1}{2}^{\pm}$ and $\frac{3}{2}^{-}$ states to initial states. The contamination to final state originating from baryon with $J^{P}=\frac{1}{2}^{-}$ is also eliminated. In Refs.~\cite{Khajouei:2025tqw,Khajouei:2024frw}, the interferences originating from the particles with $J^{P}=\frac{3}{2}^{-}$ and $J^{P}=\frac{1}{2}^{-}$ to initial and final states are neglected. Secondly, the $z$-series expansion method is used to fit the form factors in the present work, while the multipole expansion method is employed in theirs. For the transition process $\Sigma_{b}^{*0}\to\Sigma_{c}^{+}$ as an example, we explicitly shown in Fig.~\ref{zazz} the difference of the results caused by these two fitting methods. It can be seen evidently that the fitting results differ significantly in the time-like region.

\section{Semileptonic decays of $\Omega_{b}^{*}$, $\Sigma_{b}^{*}$ and $\Xi_{b}^{\prime *}$ baryons}\label{sec4}

For the weak transition process $\frac{3}{2}^{+}\rightarrow\frac{1}{2}^{+}$, the helicity amplitude can be expressed as,
\begin{equation}
H_{\lambda_2 \lambda_W}^{\mathrm{V-A}} =
\left\langle \mathcal B_{c} (p') \left| \bar{c}\gamma_\mu (1-\gamma_5) b \right| \mathcal B_{b}^* (p) \right\rangle
\epsilon^{*\mu}(\lambda_W)
\label{eq:20}
\end{equation}
where $\lambda_W = t, 0, \pm 1$ and $\lambda_2 = \pm \frac{1}{2}$ are the helicity components of the $W_ {off-shell}$ and the daughter baryon, respectively. In Eq. (\ref{eq:B1}) in Appendix~\ref{Sec:AppB}, we present the explicit relations between the helicity amplitudes \(H_{\lambda_2\lambda_W}^{\mathrm{V-A}}\) and the form factors \(F_i(q^2)\)/\(G_i(q^2)\).
The differential decay of the semileptonic decay process for $\frac{3}{2} \to \frac{1}{2}$ can be expressed as,
\begin{equation}\label{eq:21}
\frac{d\Gamma}{dq^{2}d\cos\theta}
=\frac{1}{2}\frac{G_{F}^{2}|V_{CKM}|^{2}}{192\pi^{3}}
\frac{m_{\mathcal{B}_{c}}}{m_{\mathcal{B}_{b}^{*}}^{2}}
\frac{(q^{2}-m_{l}^{2})^{2}}{q^{2}}
\sqrt{\omega^2-1}\left|M\right|^2
\end{equation}
Here $G_F = 1.16637\times10^{-5}$~GeV$^{-2}$ is the Fermi constant and $V_{cb}=0.041$ is the CKM matrix element. The kinematic variable is defined as $\omega=\frac{m_{\mathcal{B}_{b}^{*}}^{2}+m_{\mathcal{B}_{c}}^{2}-q^{2}}{2m_{\mathcal{B}_{b}^{*}}m_{\mathcal{B}_{c}}}$, with $m_{\mathcal{B}_{b}^{*}}$ and $m_{\mathcal{B}_{c}}$ being the masses of the initial and final baryons. The expression for $|M|^2$ in terms of the helicity amplitudes is given in Eq. (\ref{eq:B2}) in Appendix~\ref{Sec:AppB}. In this equation, $\theta$ is the polar angle of the lepton in the $(\ell,\nu_\ell)$ c.m. system relative to the momentum direction of the  $W_{off-shell}$.
After performing the integration over $\cos\theta$, we can obtain the differential decay width which can be decomposed into the longitudinally and transversely polarized components,
\begin{eqnarray}\label{eq:22}
\notag
&&\frac{d\Gamma_{L}}{dq^{2}}=\frac{1}{2}\frac{G_{F}^{2}|V_{CKM}|^{2}}{192\pi^{3}}\frac{m_{\mathcal{B}_{c}}}{m_{\mathcal{B}_{b}^{*}}^{2}}\frac{(q^{2}-m_{l}^{2})^{2}}{q^{2}}
\sqrt{\omega^2-1}\Bigg[|H_{\frac{1}{2},0}|^{2}\\ \notag
&&+|H_{-\frac{1}{2},0}|^{2}+\frac{m_{l}^{2}}{2q^{2}}\Big(3|H_{\frac{1}{2},t}|^{2}+3|H_{-\frac{1}{2},t}|^{2}+|H_{\frac{1}{2},0}|^{2}+|H_{-\frac{1}{2},0}|^{2}\Big)\Bigg], \\
\notag
&&\frac{d\Gamma_{T}}{dq^{2}}=\frac{1}{2}\frac{G_{F}^{2}|V_{CKM}|^{2}}{192\pi^{3}}\frac{m_{\mathcal{B}_{c}}}{m_{\mathcal{B}_{b}^{*}}^{2}}\frac{(q^{2}-m_{l}^{2})^{2}}{q^{2}}
\sqrt{\omega^2-1}\Bigg[|H_{\frac{1}{2},1}|^{2}\\ \notag
&&+|H_{-\frac{1}{2},-1}|^{2}+|H_{\frac{1}{2},-1}|^{2}+|H_{-\frac{1}{2},1}|^{2}+\frac{m_{l}^{2}}{2q^{2}}\Big(|H_{\frac{1}{2},1}|^{2}+|H_{-\frac{1}{2},-1}|^{2}\\
&&+|H_{\frac{1}{2},-1}|^{2}+|H_{-\frac{1}{2},1}|^{2}\Big)\Bigg].
\end{eqnarray}

\begin{figure*}[htbp] \centering
\includegraphics[width=1\textwidth]{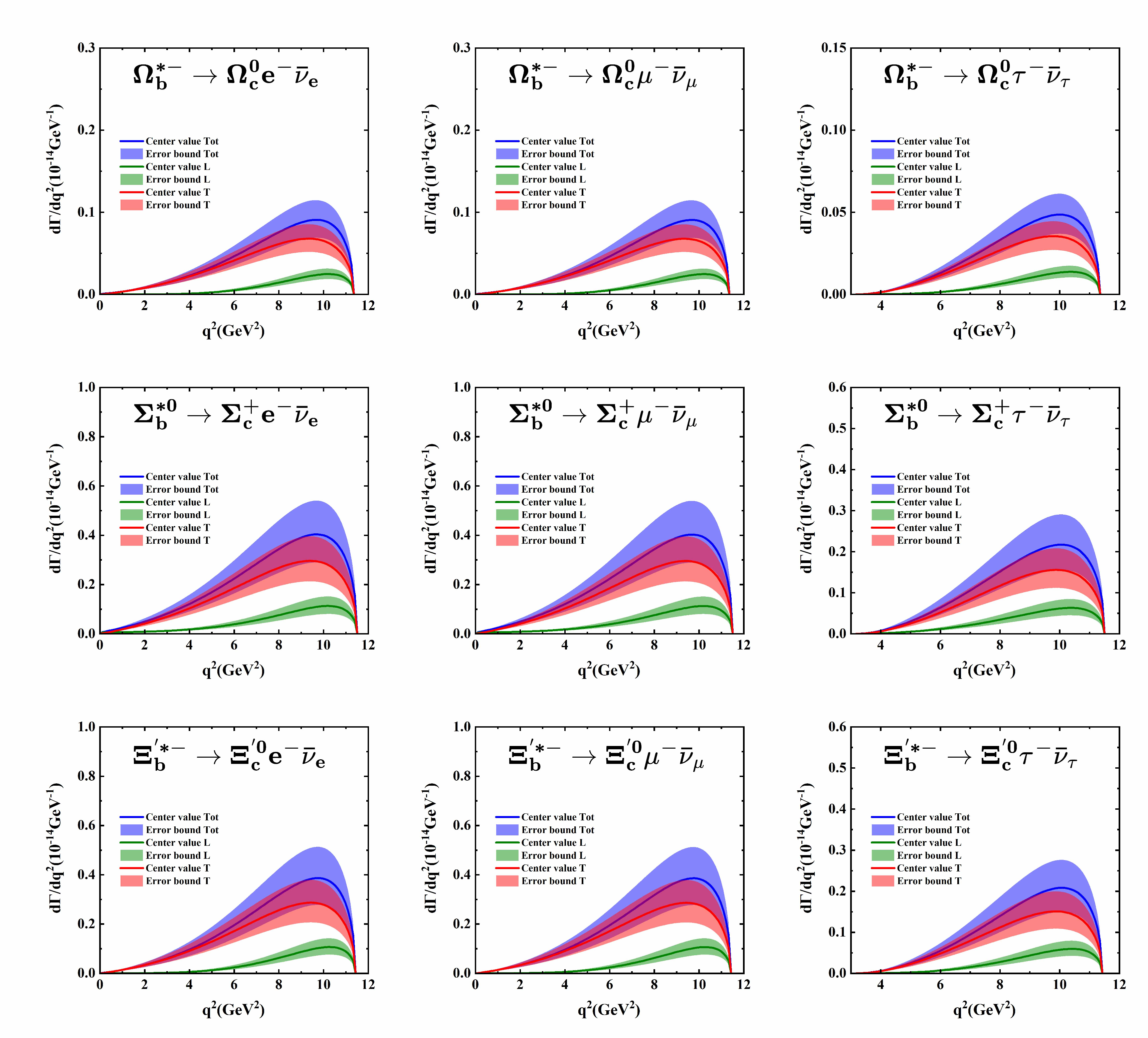}
\caption{Variations of the differential decay width with respect to $q^2$ for the semileptonic decays of $\Omega_{b}^{*-}$, $\Sigma_{b}^{*0}$ and $\Xi_{b}^{\prime*-}$ baryons.}
\label{LT}
\end{figure*}

By integrating this differential width over the allowed kinematic range $q^2 \in [m_l^2, (m_{B_b^*}-m_{B_c})^2]$, we obtain the integrated width $\Gamma$. The differential decay widths for the $\Omega_{b}^{*-}$, $\Sigma_{b}^{*0}$, and $\Xi_{b}^{\prime *-}$ baryons are explicitly shown in Fig.~\ref{LT}. The integrated widths $\Gamma$, the ratios $\Gamma_L/\Gamma_T$ and the results of other collaborations are all listed in Table~\ref{DW}. \begin{table*}[htbp]
\begin{ruledtabular}\caption{Partial widths $\Gamma$ (in units of $10^{-14}$ GeV), $\Gamma_L/\Gamma_T$, lepton forward-backward asymmetry $\langle A_{FB}^{l}\rangle$, final baryon polarization $\langle P_{z}^{F}\rangle$, and lepton polarization $\langle P_{z}^{l}\rangle$.}
\label{DW}
\begin{tabular}{c| c c c c c| c}
\multirow{2}{*}{Modes}& \multicolumn{5}{c}{This work}&\\
&$\Gamma$&$\Gamma_{L}/\Gamma_{T}$&${\langle A_{FB}^{l}\rangle}$&$\langle P_{z}^{F}\rangle$&$\langle P_{z}^{l}\rangle$&$\Gamma$ \\ \hline
$\Omega_{b}^{*-}\rightarrow\Omega_{c}^{0}e\overline{\nu}_{e}$&$2.00^{+0.53}_{-0.47}$& $0.23^{+0.008}_{-0.008}$&$-0.44^{+0.01}_{-0.01}$&$-0.67^{+0.001}_{-0.002}$&-1&$1.54^{+0.29}_{-0.27}$ \cite{Khajouei:2024frw}\\
$\Omega_{b}^{*-}\rightarrow\Omega_{c}^{0}\mu\overline{\nu}_{\mu}$&$2.00^{+0.53}_{-0.47}$& $0.23^{+0.008}_{-0.008}$&$-0.44^{+0.01}_{-0.01}$&$-0.67^{+0.001}_{-0.002}$&-1&$1.53^{+0.28}_{-0.27}$ \cite{Khajouei:2024frw}\\
$\Omega_{b}^{*-}\rightarrow\Omega_{c}^{0}\tau\overline{\nu}_{\tau}$&$0.79^{+0.21}_{-0.19}$ & $0.29^{+0.008}_{-0.008}$&$-0.33^{+0.002}_{-0.003}$&$-0.61^{+0.0001}_{-0.001}$&$-0.68^{+0.0007}_{-0.0001}$&$0.315^{+0.056}_{-0.050}$ \cite{Khajouei:2024frw}\\
$\Sigma_{b}^{*0}\rightarrow\Sigma_{c}^{+}e\overline{\nu}_{e}$&$2.37^{+0.78}_{-0.64}$& $0.30^{+0.002}_{-0.002}$&$-0.42^{+0.001}_{-0.003}$&$-0.66^{+0.0001}_{-0.003}$&-1&$134^{+38}_{-33}$ \cite{Khajouei:2025tqw}\\
$\Sigma_{b}^{*0}\rightarrow\Sigma_{c}^{+}\mu\overline{\nu}_{\mu}$&$2.36^{+0.78}_{-0.64}$ & $0.30^{+0.002}_{-0.002}$&$-0.42^{+0.001}_{-0.003}$&$-0.66^{+0.0001}_{-0.003}$&-1&$133^{+38}_{-33}$ \cite{Khajouei:2025tqw}\\
$\Sigma_{b}^{*0}\rightarrow\Sigma_{c}^{+}\tau\overline{\nu}_{\tau}$&$0.93^{+0.31}_{-0.26}$&$0.34^{+0.004}_{-0.003}$&$-0.34^{+0.001}_{-0.0003}$&$-0.60^{+0.006}_{-0.002}$&$-0.67^{+0.0005}_{-0.0006}$&$38.8^{+10.8}_{-9.5}$ \cite{Khajouei:2025tqw}\\
$\Xi_{b}^{*-}\rightarrow\Xi_{c}^{0}e\overline{\nu}_{e}$&$2.10^{+0.70}_{-0.59}$&$0.25^{+0.008}_{-0.009}$&$-0.43^{+0.005}_{-0.008}$&$-0.66^{+0.001}_{-0.005}$&-1&-\\
$\Xi_{b}^{*-}\rightarrow\Xi_{c}^{0}\mu\overline{\nu}_{\mu}$&$2.10^{+0.69}_{-0.59}$&$0.25^{+0.008}_{-0.009}$&$-0.43^{+0.005}_{-0.008}$&$-0.66^{+0.001}_{-0.005}$&-1&-\\
$\Xi_{b}^{*-}\rightarrow\Xi_{c}^{0}\tau\overline{\nu}_{\tau}$&$0.85^{+0.28}_{-0.24}$&$0.30^{+0.007}_{-0.008}$&$-0.33^{+0.001}_{-0.003}$&$-0.60^{+0.00004}_{-0.003}$&$-0.68^{+0.001}_{-0.001}$&-
\end{tabular}
\end{ruledtabular}
\end{table*}The ratios $\Gamma_L/\Gamma_T$ are in the range $0.23\sim0.30$, which indicates a dominance of the transverse polarization. By comparing with other literatures, we find that for the $\Omega_{b}^{*-} \to \Omega_{c}^{0}$ channels, our predictions are roughly comparable with those of Ref.~\cite{Khajouei:2025tqw}. For the $\Sigma_{b}^{*0} \to \Sigma_{c}^{+}$ channels, however, a significant discrepancy is observed between our results and those of Ref.~\cite{Khajouei:2024frw}. According to the SU(3) flavor symmetry, the semileptonic decay widths for these three processes
$\Omega_{b}^{*-} \to \Omega_{c}^{0} l \bar{\nu}_l$,
$\Sigma_{b}^{*0} \to \Sigma_{c}^{+} l \bar{\nu}_l$, and
$\Xi_{b}^{\prime *-} \to \Xi_{c}^{\prime 0} l \bar{\nu}_l$
should be equal to each other. As shown in Table~\ref{DW}, the predicted widths for these above three channels are roughly compatible with each other, which is consistent with the SU(3) symmetry. But the predictions are not rigourously equal to each other, for the transition processes $\Omega_{b}^{*-} \to \Omega_{c}^{0} e \bar{\nu}_e$,
$\Sigma_{b}^{*0} \to \Sigma_{c}^{+} e \bar{\nu}_e$, and
$\Xi_{b}^{\prime *-} \to \Xi_{c}^{\prime 0} e \bar{\nu}_e$ as examples, the central values of our predictions are $2.00\times10^{-14}$, $2.37\times10^{-14}$ and $2.10\times10^{-14}$ GeV, respectively. This difference reflects the SU(3) symmetry breaking effect which can be
explained by the input parameters in the present work. The masses of the $u$ and $d$ quarks are neglected, while the mass of strange quark is determined by the RGE in Eq. (\ref{eq:17}). We also note from Refs.~\cite{Khajouei:2025tqw,Khajouei:2024frw} that their predicted width for $\Sigma_{b}^{*0} \to \Sigma_{c}^{+} l \bar{\nu}_l$ channel is two orders of magnitude larger than that of $\Omega_{b}^{*-} \to \Omega_{c}^{0} l \bar{\nu}_l$ transition process. This reflect that the SU(3) flavor symmetry is significantly broken in these semileptonic decays. Thus, more theoretical analyses and precise experimental measurements are crucial in the future to confirm or figure out this problem.

Besides of the decay widths, several physical observables for the semileptonic process can be analyzed  \cite{Kadeer:2005aq}, including the forward-backward asymmetry $A_{\rm FB}^l$ of the lepton, the polarization component $P_z^F$ of the final-state baryon $\mathcal B_c$, and the longitudinal lepton polarization $P_z^l$ for an unpolarized initial baryon. These observables are expressed as,
\begin{eqnarray}\label{eq:23}
\notag
&&A_{FB}^{l}(q^2)\\ \notag
&&=-\frac{3}{4}
\Bigg\{\dfrac{\left(|H_{\frac{1}{2}1}|^2+|H_{-\frac{1}{2}1}|^2\right)-\left(|H_{-\frac{1}{2}-1}|^2+|H_{\frac{1}{2}-1}|^2\right)}{d\Gamma/dq^2}\\
&&-\dfrac{2m_l^2}{q^2}\dfrac{H_{\frac{1}{2}t}^*H_{\frac{1}{2}0}+H_{-\frac{1}{2}t}^*H_{-\frac{1}{2}0}}{d\Gamma/dq^2}\Bigg\}
\end{eqnarray}
\begin{eqnarray}\label{eq:24}
P_z^{F}(\theta)=\frac{L_{\mu\nu}H_{++}^{\mu\nu}-L_{\mu\nu}H_{--}^{\mu\nu}}{L_{\mu\nu}H_{++}^{\mu\nu}+L_{\mu\nu}H_{--}^{\mu\nu}}
\end{eqnarray}
\begin{eqnarray}\label{eq:25}
P_z^{l}(\theta)=\frac{L_{\mu\nu}H^{\mu\nu}(flip)-L_{\mu\nu}H^{\mu\nu}(nonflip)}{L_{\mu\nu}H^{\mu\nu}(flip)+L_{\mu\nu}H^{\mu\nu}(nonflip)}
\end{eqnarray}
with
\begin{flalign}
\begin{aligned}
&L_{\mu\nu}H_{--}^{\mu\nu} = \frac{2}{3}(q^2-m_l^2)
\Bigg\{\frac{3}{8}(1+\cos\theta)^2\left(|H_{-\frac{1}{2}-1}|^2+|H_{\frac{1}{2}-1}|^2\right) \\
&\quad +\frac{3}{4}\sin^2\theta|H_{-\frac{1}{2}0}|^2+\frac{m_l^2}{2q^2}\Bigg[\frac{3}{2}|H_{-\frac{1}{2}t}|^2
+\frac{3}{4}\sin^2\theta\left(|H_{-\frac{1}{2}-1}|^2\right. \\
&\quad \left.+|H_{\frac{1}{2}-1}|^2\right)+\frac{3}{2}\cos^2\theta|H_{-\frac{1}{2}0}|^2-3\cos\theta H_{-\frac{1}{2}t}H_{-\frac{1}{2}0}
\Bigg]\Bigg\}
\end{aligned} \label{eq:26}
\end{flalign}
\begin{flalign}
\begin{aligned}
&L_{\mu\nu}H_{++}^{\mu\nu} = \frac{2}{3}(q^2-m_l^2)
\Bigg\{\frac{3}{8}(1-\cos\theta)^2\left(|H_{\frac{1}{2}1}|^2+|H_{-\frac{1}{2}1}|^2\right) \\
&\quad +\frac{3}{4}\sin^2\theta|H_{\frac{1}{2}0}|^2+\frac{m_l^2}{2q^2}\Bigg[\frac{3}{2}|H_{\frac{1}{2}t}|^2
+\frac{3}{4}\sin^2\theta\left(|H_{\frac{1}{2}1}|^2\right. \\
&\quad \left.+|H_{-\frac{1}{2}1}|^2\right)+\frac{3}{2}\cos^2\theta|H_{\frac{1}{2}0}|^2-3\cos\theta H_{\frac{1}{2}t}H_{\frac{1}{2}0}
\Bigg]\Bigg\}
\end{aligned} \label{eq:27}
\end{flalign}
\begin{figure*}[htbp] \centering
\includegraphics[width=1\textwidth]{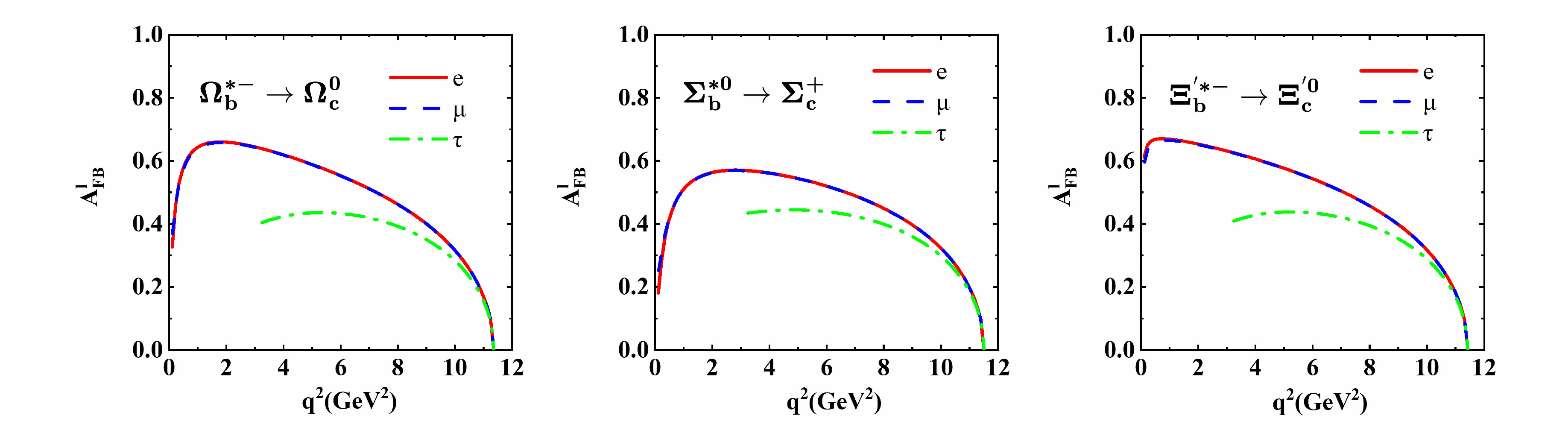}
\caption{Dependence of forward-backward asymmetry $A_{FB}^{l}$ of the lepton on $q^{2}$.}
\label{AFB}
\centering
\includegraphics[width=1\textwidth]{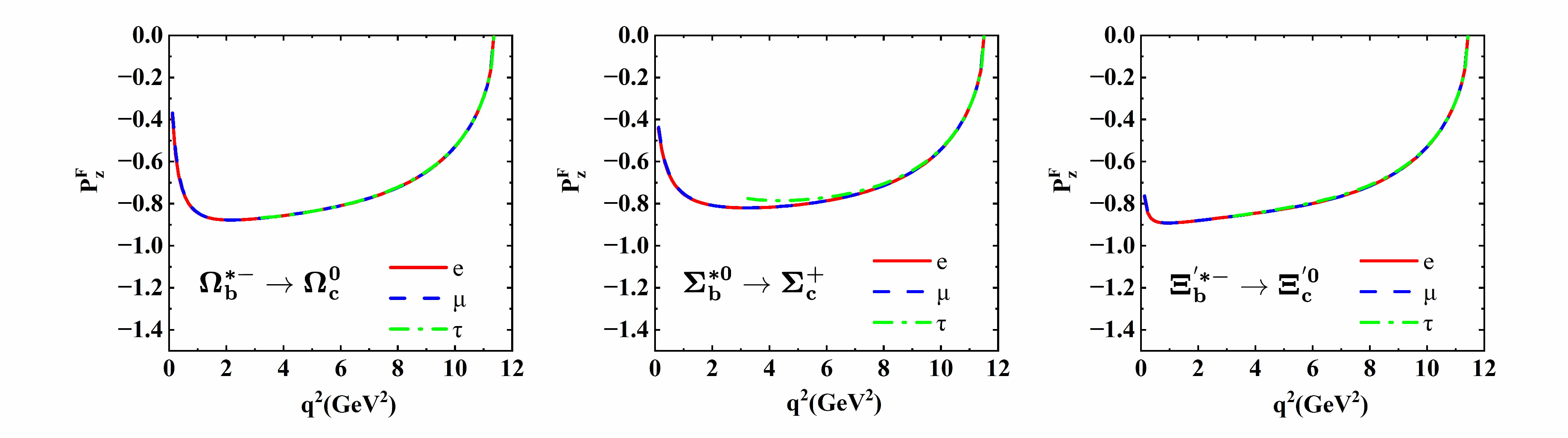}
\caption{Same as Fig.~\ref{AFB} but for the component $P_z^{F}(\theta)$ of the polarization vector of daughter baryon.}
\label{PZF}
\centering
\includegraphics[width=1\textwidth]{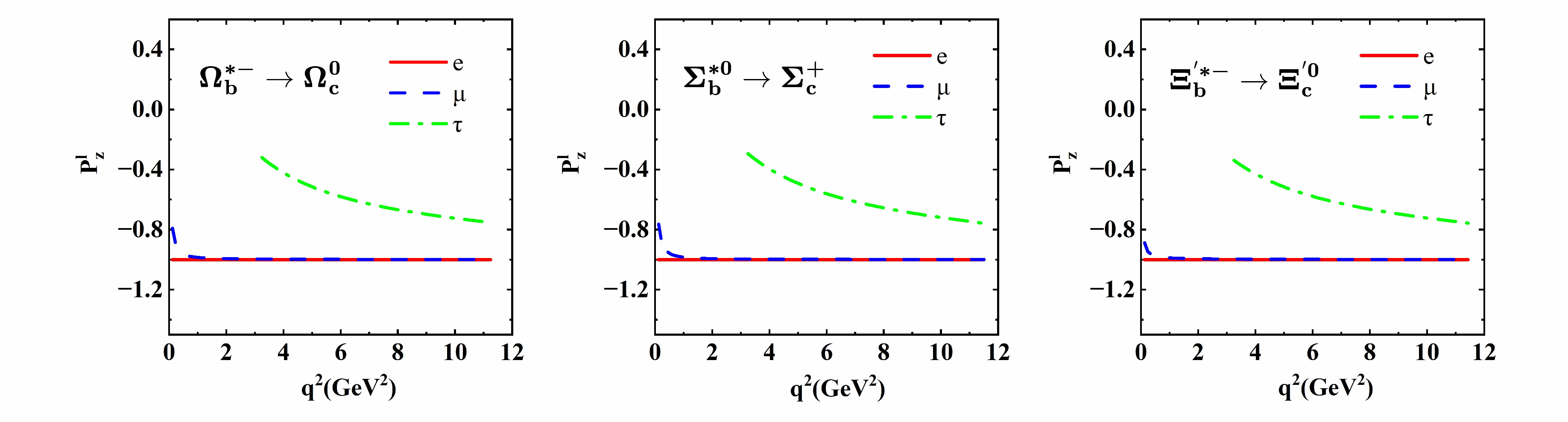}
\caption{Same as Fig.~\ref{AFB} but for the longitudinal polarization $P_z^{l}$ of the lepton.}
\label{PZL}
\end{figure*}
Finishing the integration of $\cos\theta$ for $P_z^{F}(\theta)$ and $P_z^{l}(\theta)$, we obtain the momentum dependent $P_z^F(q^2)$ and $P_z^l(q^2)$. In Figs.~\ref{AFB}$\sim$~\ref{PZL}, we illustrate the dependence of $A_{\rm FB}^l(q^2)$, $P_z^F(q^2)$, and $P_z^l(q^2)$ on $q^2$. From these figures, we can see that the $\tau$ decaying mode is evidently different with $e$ and $\mu$ channels for the forward-backward asymmetry and the longitudinal polarization of the lepton (See Figs. \ref{AFB} and \ref{PZL}). This is due to the much larger mass of $\tau$ than $e$ and $\mu$. Finally, after performing the integration over $q^2$ in the range $m_l^2 \le q^2 \le (m_{\mathcal B_{b}^{*}} - m_{\mathcal B_{c}})^2$, we obtain the averaged values of $\langle A_{\rm FB}^l \rangle$, $\langle P_z^F \rangle$, and $\langle P_z^l \rangle$, which are also listed in Table~\ref{DW}. The values of $\langle A_{\rm FB}^l \rangle$ for the $e$ and $\mu$ modes are approximately in the range $-0.44 \sim -0.42$, while for the $\tau$ mode, they lie in the range $-0.34 \sim -0.33$. The predicted values of $\langle P_z^F \rangle$ are close to each other for the three baryon transitions, with values around $-0.67 \sim -0.60$. As for the longitudinal polarization $\langle P_z^l \rangle$, the flip transition part in Eq. (\ref{eq:25}) is related to the factor of $\frac{m_{l}^{2}}{2q^{2}}$ (See Eq. (\ref{eq:B2})). Thus, the values of $\langle P_z^l \rangle$ for $e$ and $\mu$ modes are approximately $-1$, because the masses of these leptons are close to zero. While for the $\tau$ decaying mode, their values vary in the range $-0.67$ and $-0.68$.

\section{Conclusions}\label{sec5}

In the present work, we firstly analyze the transition form factors for the semileptonic decays
$\Sigma_b^{*0} \to \Sigma_c^{+} l\bar{\nu}_l$,
$\Xi_b^{\prime*-} \to \Xi_c^{\prime0} l\bar{\nu}_l$, and
$\Omega_b^{*-} \to \Omega_c^{0} l\bar{\nu}_l$.
In this analysis, we consider the couplings of the currents to the baryons with low spin and negative parity, and eliminate the contaminations of these couplings. By employing all of the dirac structures, 16 linear equations about the form factors are constructed to extract the results for the transition process $\frac{3}{2}^{+}\rightarrow\frac{1}{2}^{+}$.
With the predicted form factors, we finally calculate the decay widths and find that SU(3) flavor symmetry is mildly broken among the three channels. In addition, we present predictions for the forward-backward asymmetry $A_{FB}$, the final-state baryon polarization $P_z^F$, and the lepton longitudinal polarization $P_z^l$. These observables are sensitive to possible new physics beyond the Standard Model. We expect that the predictions in the present work can provide a valuable reference for both theoretical and experimental studies of weak decays involving heavy flavor baryons.

\section*{Acknowledgements}
This project is supported by National Natural Science
Foundation,Grant Number 12175068 and Natural Science
Foundation of HeBei Province,Grant Number A2024502002.

\appendix
\onecolumngrid
\section{QCD sum rules for the form factors}\label{Sec:AppA}

\begin{eqnarray*}
F_{1}^{++}(Q^2) & = & -\frac{\text{exp}\big[\frac{m_{\mathcal{B}_{b}^{*+}}^{2}}{M_{1}^{2}}+\frac{m_{\mathcal{B}_{c}^{+}}^{2}}{M_{2}^{2}}\big]}
{4\lambda_{\mathcal{B}_{b}^{*+}}\lambda_{\mathcal{B}_{c}^{+}}m_{\mathcal{B}_{b}^{*+}}\Bigl(m_{\mathcal{B}_{c}^{+}}+m_{\mathcal{B}_{c}^{-}}\Bigr)}
\int^{s_{0}}_{s_{min}}ds\int^{u_{0}}_{u_{min}}du\,\text{exp}\big[-\frac{s}{M_1^2}-\frac{u}{M_2^2}\big]
\Big\{ m_{\mathcal{B}_{b}^{*+}}\rho_{4}^{\mathrm{QCD-V}}
\Bigl(m_{\mathcal{B}_{b}^{*+}}^2+2 m_{\mathcal{B}_{b}^{*+}} m_{\mathcal{B}_{c}^{-}}\\
 & &+m_{\mathcal{B}_{b}^{*-}}^2+2 \bigl(m_{\mathcal{B}_{c}^{+}} m_{\mathcal{B}_{c}^{-}}+Q^2\bigr)\Bigr) + \rho_{3}^{\mathrm{QCD-V}}
\Bigl(m_{\mathcal{B}_{b}^{*+}}^2+2 m_{\mathcal{B}_{b}^{*+}} m_{\mathcal{B}_{c}^{-}}-m_{\mathcal{B}_{b}^{*-}}^2\Bigr)\\
 & &
-2 m_{\mathcal{B}_{b}^{*+}} \rho_{2}^{\mathrm{QCD-V}} (m_{\mathcal{B}_{b}^{*+}}+m_{\mathcal{B}_{c}^{+}}-m_{\mathcal{B}_{c}^{-}})
+2 m_{\mathcal{B}_{b}^{*+}} \rho_{1}^{\mathrm{QCD-V}}
\Big\} \\
F_{2}^{++}(Q^2) & = & -\frac{\text{exp}\big[\frac{m_{\mathcal{B}_{b}^{*+}}^{2}}{M_{1}^{2}}+\frac{m_{\mathcal{B}_{c}^{+}}^{2}}{M_{2}^{2}}\big]}
{2 \lambda_{\mathcal{B}_{b}^{*+}} \lambda_{\mathcal{B}_{c}^{+}} m_{\mathcal{B}_{b}^{*+}}
(m_{\mathcal{B}_{c}^{+}}+m_{\mathcal{B}_{c}^{-}}) \Bigl(2 m_{\mathcal{B}_{b}^{*+}}^2 m_{\mathcal{B}_{b}^{*-}}+3 m_{\mathcal{B}_{b}^{*+}}
(m_{\mathcal{B}_{c}^{+}}-m_{\mathcal{B}_{c}^{-}})^2+m_{\mathcal{B}_{b}^{*-}} (m_{\mathcal{B}_{c}^{+}}-m_{\mathcal{B}_{c}^{-}})^2\Bigr)}\\
 & &
\int^{s_{0}}_{s_{min}}ds\int^{u_{0}}_{u_{min}}du\,\text{exp}\big[-\frac{s}{M_1^2}-\frac{u}{M_2^2}\big]
 m_{\mathcal{B}_{c}^{+}} \Big( -4 m_{\mathcal{B}_{b}^{*+}}^2 \rho_{3}^{\mathrm{QCD-V}} (m_{\mathcal{B}_{b}^{*-}}+m_{\mathcal{B}_{c}^{+}}-m_{\mathcal{B}_{c}^{-}})-4 m_{\mathcal{B}_{b}^{*+}}^2 \rho_{4}^{\mathrm{QCD-V}} \bigl(m_{\mathcal{B}_{b}^{*+}} (m_{\mathcal{B}_{b}^{*-}}\\
 & &-m_{\mathcal{B}_{c}^{+}}+m_{\mathcal{B}_{c}^{-}})+m_{\mathcal{B}_{b}^{*-}} (m_{\mathcal{B}_{c}^{-}}-m_{\mathcal{B}_{c}^{+}})\bigr) +6 m_{\mathcal{B}_{b}^{*+}}^2 \rho_{5}^{\mathrm{QCD-V}} (m_{\mathcal{B}_{b}^{*-}}+m_{\mathcal{B}_{c}^{+}}-m_{\mathcal{B}_{c}^{-}})
-6 m_{\mathcal{B}_{b}^{*+}}^2 \rho_{6}^{\mathrm{QCD-V}} \bigl(m_{\mathcal{B}_{b}^{*+}} (m_{\mathcal{B}_{b}^{*-}}\\
 & &-m_{\mathcal{B}_{c}^{+}}+m_{\mathcal{B}_{c}^{-}})+m_{\mathcal{B}_{b}^{*-}} (m_{\mathcal{B}_{c}^{-}}-m_{\mathcal{B}_{c}^{+}})\bigr) + \rho_{7}^{\mathrm{QCD-V}} \Bigl( m_{\mathcal{B}_{b}^{*+}}^3 (m_{\mathcal{B}_{b}^{*-}}+3 m_{\mathcal{B}_{c}^{+}}-3 m_{\mathcal{B}_{c}^{-}})
- m_{\mathcal{B}_{b}^{*+}}^2 \bigl(2 m_{\mathcal{B}_{b}^{*-}}^2-m_{\mathcal{B}_{b}^{*-}} m_{\mathcal{B}_{c}^{+}}\\
 & &+3 m_{\mathcal{B}_{b}^{*-}} m_{\mathcal{B}_{c}^{-}}+2 m_{\mathcal{B}_{c}^{+}} m_{\mathcal{B}_{c}^{-}}+2 Q^2\bigr) - m_{\mathcal{B}_{b}^{*+}} m_{\mathcal{B}_{b}^{*-}}^3
+3 m_{\mathcal{B}_{b}^{*+}} (m_{\mathcal{B}_{c}^{+}}-m_{\mathcal{B}_{c}^{-}}) \bigl(m_{\mathcal{B}_{c}^{-}}^2+Q^2\bigr)
+m_{\mathcal{B}_{b}^{*-}} (m_{\mathcal{B}_{c}^{+}}-m_{\mathcal{B}_{c}^{-}}) \bigl(m_{\mathcal{B}_{c}^{-}}^2\\
 & &+Q^2\bigr) \Bigr) + m_{\mathcal{B}_{b}^{*+}} \rho_{8}^{\mathrm{QCD-V}}  \Bigl( m_{\mathcal{B}_{b}^{*+}}^3 (m_{\mathcal{B}_{b}^{*-}}+3 m_{\mathcal{B}_{c}^{+}}-3 m_{\mathcal{B}_{c}^{-}})+ m_{\mathcal{B}_{b}^{*+}}^2 \bigl(2 m_{\mathcal{B}_{b}^{*-}}^2+m_{\mathcal{B}_{b}^{*-}} (m_{\mathcal{B}_{c}^{+}}-3 m_{\mathcal{B}_{c}^{-}})+2 \bigl(m_{\mathcal{B}_{c}^{+}} m_{\mathcal{B}_{c}^{-}}+Q^2\bigr)\bigr) \\
 & & + m_{\mathcal{B}_{b}^{*+}} \bigl(m_{\mathcal{B}_{b}^{*-}}^3+2 m_{\mathcal{B}_{b}^{*-}} \bigl(m_{\mathcal{B}_{c}^{+}} m_{\mathcal{B}_{c}^{-}}+Q^2\bigr)+3 (m_{\mathcal{B}_{c}^{+}}-m_{\mathcal{B}_{c}^{-}}) \bigl(m_{\mathcal{B}_{c}^{-}}^2+Q^2\bigr)\bigr) + m_{\mathcal{B}_{b}^{*-}} (m_{\mathcal{B}_{c}^{+}}-m_{\mathcal{B}_{c}^{-}}) \bigl(m_{\mathcal{B}_{c}^{-}}^2+Q^2\bigr) \Bigr) \Big) \\
F_{3}^{++}(Q^2) & = & \frac{\text{exp}\big[\frac{m_{\mathcal{B}_{b}^{*+}}^{2}}{M_{1}^{2}}+\frac{m_{\mathcal{B}_{c}^{+}}^{2}}{M_{2}^{2}}\big]}
{2 \lambda_{\mathcal{B}_{b}^{*+}} \lambda_{\mathcal{B}_{c}^{+}}
(m_{\mathcal{B}_{c}^{+}}+m_{\mathcal{B}_{c}^{-}}) \Bigl(2 m_{\mathcal{B}_{b}^{*+}}^2 m_{\mathcal{B}_{b}^{*-}}+3 m_{\mathcal{B}_{b}^{*+}}
(m_{\mathcal{B}_{c}^{+}}-m_{\mathcal{B}_{c}^{-}})^2+m_{\mathcal{B}_{b}^{*-}} (m_{\mathcal{B}_{c}^{+}}-m_{\mathcal{B}_{c}^{-}})^2\Bigr)}\\
 & &
\int^{s_{0}}_{s_{min}}ds\int^{u_{0}}_{u_{min}}du\,\text{exp}\big[-\frac{s}{M_1^2}-\frac{u}{M_2^2}\big]
\Big( 6 m_{\mathcal{B}_{b}^{*+}}^2 m_{\mathcal{B}_{c}^{+}} \rho_{10}^{\mathrm{QCD-V}}
(-m_{\mathcal{B}_{c}^{-}} (m_{\mathcal{B}_{b}^{*+}}+m_{\mathcal{B}_{b}^{*-}})+m_{\mathcal{B}_{b}^{*+}} m_{\mathcal{B}_{b}^{*-}}+m_{\mathcal{B}_{b}^{*+}} m_{\mathcal{B}_{c}^{+}}\\
 & &+m_{\mathcal{B}_{b}^{*-}} m_{\mathcal{B}_{c}^{+}})  + 2 m_{\mathcal{B}_{c}^{+}} \rho_{3}^{\mathrm{QCD-V}} \bigl(2 m_{\mathcal{B}_{b}^{*+}}^2-m_{\mathcal{B}_{c}^{-}} (3 m_{\mathcal{B}_{b}^{*+}}+m_{\mathcal{B}_{b}^{*-}})+3 m_{\mathcal{B}_{b}^{*+}} m_{\mathcal{B}_{c}^{+}}+m_{\mathcal{B}_{b}^{*-}} m_{\mathcal{B}_{c}^{+}}\bigr) \\
 & & + 2 m_{\mathcal{B}_{b}^{*+}} m_{\mathcal{B}_{c}^{+}} \rho_{4}^{\mathrm{QCD-V}} \bigl(-2 m_{\mathcal{B}_{b}^{*+}}^2+m_{\mathcal{B}_{b}^{*+}} (4 m_{\mathcal{B}_{b}^{*-}}-3 m_{\mathcal{B}_{c}^{+}}+3 m_{\mathcal{B}_{c}^{-}})+m_{\mathcal{B}_{b}^{*-}} (m_{\mathcal{B}_{c}^{+}}-m_{\mathcal{B}_{c}^{-}})\bigr) \\
 & & -6 m_{\mathcal{B}_{b}^{*+}}^2 m_{\mathcal{B}_{c}^{+}} \rho_{9}^{\mathrm{QCD-V}} (m_{\mathcal{B}_{b}^{*-}}-m_{\mathcal{B}_{c}^{+}}+m_{\mathcal{B}_{c}^{-}}) + m_{\mathcal{B}_{c}^{+}} \rho_{11}^{\mathrm{QCD-V}} \Bigl( -m_{\mathcal{B}_{b}^{*+}}^3 (m_{\mathcal{B}_{b}^{*-}}-3 m_{\mathcal{B}_{c}^{+}}+3 m_{\mathcal{B}_{c}^{-}}) \\
 & & + m_{\mathcal{B}_{b}^{*+}}^2 \bigl(2 m_{\mathcal{B}_{b}^{*-}}^2+m_{\mathcal{B}_{b}^{*-}} (m_{\mathcal{B}_{c}^{+}}-3 m_{\mathcal{B}_{c}^{-}})+2 \bigl(m_{\mathcal{B}_{c}^{+}} m_{\mathcal{B}_{c}^{-}}+Q^2\bigr)\bigr) + m_{\mathcal{B}_{b}^{*+}} \bigl(m_{\mathcal{B}_{b}^{*-}}^3+3 (m_{\mathcal{B}_{c}^{+}}-m_{\mathcal{B}_{c}^{-}}) \bigl(m_{\mathcal{B}_{c}^{-}}^2+Q^2\bigr)\bigr) \\
 & & + m_{\mathcal{B}_{b}^{*-}} (m_{\mathcal{B}_{c}^{+}}-m_{\mathcal{B}_{c}^{-}}) \bigl(m_{\mathcal{B}_{c}^{-}}^2+Q^2\bigr) \Bigr) - m_{\mathcal{B}_{b}^{*+}} m_{\mathcal{B}_{c}^{+}} \rho_{12}^{\mathrm{QCD-V}} \Bigl( m_{\mathcal{B}_{b}^{*+}}^3 (m_{\mathcal{B}_{b}^{*-}}-3 m_{\mathcal{B}_{c}^{+}}+3 m_{\mathcal{B}_{c}^{-}}) \\
 & & + m_{\mathcal{B}_{b}^{*+}}^2 \bigl(2 m_{\mathcal{B}_{b}^{*-}}^2-m_{\mathcal{B}_{b}^{*-}} m_{\mathcal{B}_{c}^{+}}+3 m_{\mathcal{B}_{b}^{*-}} m_{\mathcal{B}_{c}^{-}}+2 m_{\mathcal{B}_{c}^{+}} m_{\mathcal{B}_{c}^{-}}+2 Q^2\bigr)  + m_{\mathcal{B}_{b}^{*+}} \bigl(m_{\mathcal{B}_{b}^{*-}}^3+2 m_{\mathcal{B}_{b}^{*-}} \bigl(m_{\mathcal{B}_{c}^{+}} m_{\mathcal{B}_{c}^{-}}+Q^2\bigr)\\
 & &-3 (m_{\mathcal{B}_{c}^{+}}-m_{\mathcal{B}_{c}^{-}}) \bigl(m_{\mathcal{B}_{c}^{-}}^2+Q^2\bigr)\bigr) - m_{\mathcal{B}_{b}^{*-}} (m_{\mathcal{B}_{c}^{+}}-m_{\mathcal{B}_{c}^{-}}) \bigl(m_{\mathcal{B}_{c}^{-}}^2+Q^2\bigr) \Bigr) \Big) \\
F_{4}^{++}(Q^2) & = & \frac{\text{exp}\big[\frac{m_{\mathcal{B}_{b}^{*+}}^{2}}{M_{1}^{2}}+\frac{m_{\mathcal{B}_{c}^{+}}^{2}}{M_{2}^{2}}\big]}
{2 \lambda_{\mathcal{B}_{b}^{*+}} \lambda_{\mathcal{B}_{c}^{+}} m_{\mathcal{B}_{b}^{*+}}
(m_{\mathcal{B}_{c}^{+}}+m_{\mathcal{B}_{c}^{-}}) \Bigl(2 m_{\mathcal{B}_{b}^{*+}}^2 m_{\mathcal{B}_{b}^{*-}}+3 m_{\mathcal{B}_{b}^{*+}}
(m_{\mathcal{B}_{c}^{+}}-m_{\mathcal{B}_{c}^{-}})^2+m_{\mathcal{B}_{b}^{*-}} (m_{\mathcal{B}_{c}^{+}}-m_{\mathcal{B}_{c}^{-}})^2\Bigr)}\\
 & &
\int^{s_{0}}_{s_{min}}ds\int^{u_{0}}_{u_{min}}du\,\text{exp}\big[-\frac{s}{M_1^2}-\frac{u}{M_2^2}\big]
\Big( -6 m_{\mathcal{B}_{b}^{*+}}^2 m_{\mathcal{B}_{c}^{+}}^2 \rho_{13}^{\mathrm{QCD-V}} (m_{\mathcal{B}_{b}^{*-}}-m_{\mathcal{B}_{c}^{+}}+m_{\mathcal{B}_{c}^{-}}) \\
 & &+ 6 m_{\mathcal{B}_{b}^{*+}}^2 m_{\mathcal{B}_{c}^{+}}^2 \rho_{14}^{\mathrm{QCD-V}} (-m_{\mathcal{B}_{c}^{-}} (m_{\mathcal{B}_{b}^{*+}}+m_{\mathcal{B}_{b}^{*-}})+m_{\mathcal{B}_{b}^{*+}} m_{\mathcal{B}_{b}^{*-}}+m_{\mathcal{B}_{b}^{*+}} m_{\mathcal{B}_{c}^{+}}+m_{\mathcal{B}_{b}^{*-}} m_{\mathcal{B}_{c}^{+}}) \\
 & & -4 m_{\mathcal{B}_{b}^{*+}}^2 m_{\mathcal{B}_{c}^{+}}^2 \rho_{7}^{\mathrm{QCD-V}} (m_{\mathcal{B}_{b}^{*-}}-m_{\mathcal{B}_{c}^{+}}+m_{\mathcal{B}_{c}^{-}})
-4 m_{\mathcal{B}_{b}^{*+}}^2 m_{\mathcal{B}_{c}^{+}}^2 \rho_{8}^{\mathrm{QCD-V}} (-m_{\mathcal{B}_{c}^{-}} (m_{\mathcal{B}_{b}^{*+}}+m_{\mathcal{B}_{b}^{*-}})+m_{\mathcal{B}_{b}^{*+}} m_{\mathcal{B}_{b}^{*-}}\\
 & &+m_{\mathcal{B}_{b}^{*+}} m_{\mathcal{B}_{c}^{+}}+m_{\mathcal{B}_{b}^{*-}} m_{\mathcal{B}_{c}^{+}})  + m_{\mathcal{B}_{c}^{+}}^2 \rho_{15}^{\mathrm{QCD-V}} \Bigl( -m_{\mathcal{B}_{b}^{*+}}^3 (m_{\mathcal{B}_{b}^{*-}}-3 m_{\mathcal{B}_{c}^{+}}+3 m_{\mathcal{B}_{c}^{-}}) + m_{\mathcal{B}_{b}^{*+}}^2 \bigl(2 m_{\mathcal{B}_{b}^{*-}}^2+m_{\mathcal{B}_{b}^{*-}} (m_{\mathcal{B}_{c}^{+}}-3 m_{\mathcal{B}_{c}^{-}})\\
 & &+2 \bigl(m_{\mathcal{B}_{c}^{+}} m_{\mathcal{B}_{c}^{-}}+Q^2\bigr)\bigr)  + m_{\mathcal{B}_{b}^{*+}} \bigl(m_{\mathcal{B}_{b}^{*-}}^3+3 (m_{\mathcal{B}_{c}^{+}}-m_{\mathcal{B}_{c}^{-}}) \bigl(m_{\mathcal{B}_{c}^{-}}^2+Q^2\bigr)\bigr) + m_{\mathcal{B}_{b}^{*-}} (m_{\mathcal{B}_{c}^{+}}-m_{\mathcal{B}_{c}^{-}}) \bigl(m_{\mathcal{B}_{c}^{-}}^2+Q^2\bigr) \Bigr) \\
 & &- m_{\mathcal{B}_{b}^{*+}} m_{\mathcal{B}_{c}^{+}}^2 \rho_{16}^{\mathrm{QCD-V}} \Bigl( m_{\mathcal{B}_{b}^{*+}}^3 (m_{\mathcal{B}_{b}^{*-}}-3 m_{\mathcal{B}_{c}^{+}}+3 m_{\mathcal{B}_{c}^{-}}) + m_{\mathcal{B}_{b}^{*+}}^2 \bigl(2 m_{\mathcal{B}_{b}^{*-}}^2-m_{\mathcal{B}_{b}^{*-}} m_{\mathcal{B}_{c}^{+}}+3 m_{\mathcal{B}_{b}^{*-}} m_{\mathcal{B}_{c}^{-}}+2 m_{\mathcal{B}_{c}^{+}} m_{\mathcal{B}_{c}^{-}}+2 Q^2\bigr) \\
 & & + m_{\mathcal{B}_{b}^{*+}} \bigl(m_{\mathcal{B}_{b}^{*-}}^3+2 m_{\mathcal{B}_{b}^{*-}} \bigl(m_{\mathcal{B}_{c}^{+}} m_{\mathcal{B}_{c}^{-}}+Q^2\bigr)-3 (m_{\mathcal{B}_{c}^{+}}-m_{\mathcal{B}_{c}^{-}}) \bigl(m_{\mathcal{B}_{c}^{-}}^2+Q^2\bigr)\bigr) - m_{\mathcal{B}_{b}^{*-}} (m_{\mathcal{B}_{c}^{+}}-m_{\mathcal{B}_{c}^{-}}) \bigl(m_{\mathcal{B}_{c}^{-}}^2+Q^2\bigr) \Bigr) \Big)
\end{eqnarray*}

\begin{eqnarray*}
G_{1}^{++}(Q^2) & = & \frac{\text{exp}\big[\frac{m_{\mathcal{B}_{b}^{*+}}^{2}}{M_{1}^{2}}+\frac{m_{\mathcal{B}_{c}^{+}}^{2}}{M_{2}^{2}}\big]}
{\lambda_{\mathcal{B}_{b}^{*+}}\lambda_{\mathcal{B}_{c}^{+}} (m_{\mathcal{B}_{b}^{*+}}+m_{\mathcal{B}_{b}^{*-}}) (m_{\mathcal{B}_{c}^{+}}+m_{\mathcal{B}_{c}^{-}})}
\int^{s_{0}}_{s_{min}}ds\int^{u_{0}}_{u_{min}}du\,\text{exp}\big[-\frac{s}{M_1^2}-\frac{u}{M_2^2}\big]
\Big\{ \rho_{1}^{\mathrm{QCD-V}} \bigl(m_{\mathcal{B}_{b}^{*+}} m_{\mathcal{B}_{b}^{*-}}-m_{\mathcal{B}_{b}^{*-}} m_{\mathcal{B}_{c}^{-}}\\
 & &+m_{\mathcal{B}_{c}^{+}} m_{\mathcal{B}_{c}^{-}}+Q^2\bigr) + \rho_{3}^{\mathrm{QCD-V}} (-m_{\mathcal{B}_{b}^{*+}}+m_{\mathcal{B}_{b}^{*-}}+m_{\mathcal{B}_{c}^{-}})
+ \rho_{2}^{\mathrm{QCD-V}} (m_{\mathcal{B}_{b}^{*-}}-m_{\mathcal{B}_{c}^{+}}+m_{\mathcal{B}_{c}^{-}})
+ \rho_{4}^{\mathrm{QCD-V}} \Big\} \\
G_{2}^{++}(Q^2) & = & \frac{\text{exp}\big[\frac{m_{\mathcal{B}_{b}^{*+}}^{2}}{M_{1}^{2}}+\frac{m_{\mathcal{B}_{c}^{+}}^{2}}{M_{2}^{2}}\big]}
{\lambda_{\mathcal{B}_{b}^{*+}}\lambda_{\mathcal{B}_{c}^{+}} (m_{\mathcal{B}_{b}^{*+}}+m_{\mathcal{B}_{b}^{*-}}) (m_{\mathcal{B}_{c}^{+}}+m_{\mathcal{B}_{c}^{-}}) \bigl(m_{\mathcal{B}_{b}^{*+}} (m_{\mathcal{B}_{b}^{*-}}-m_{\mathcal{B}_{c}^{+}}+m_{\mathcal{B}_{c}^{-}})+(m_{\mathcal{B}_{c}^{+}}-m_{\mathcal{B}_{c}^{-}}) (m_{\mathcal{B}_{b}^{*-}}+2 m_{\mathcal{B}_{c}^{+}}-2 m_{\mathcal{B}_{c}^{-}})\bigr)}\\
 & &
\int^{s_{0}}_{s_{min}}ds\int^{u_{0}}_{u_{min}}du\,\text{exp}\big[-\frac{s}{M_1^2}-\frac{u}{M_2^2}\big]
\Big\{ \rho_{7}^{\mathrm{QCD-V}} \Bigg(2 m_{\mathcal{B}_{b}^{*+}}^2 (m_{\mathcal{B}_{b}^{*-}}-m_{\mathcal{B}_{c}^{+}}+m_{\mathcal{B}_{c}^{-}})+m_{\mathcal{B}_{b}^{*+}} \bigl(-3 m_{\mathcal{B}_{b}^{*-}}^2+m_{\mathcal{B}_{b}^{*-}} m_{\mathcal{B}_{c}^{-}}\\
 & &+m_{\mathcal{B}_{c}^{-}} (m_{\mathcal{B}_{c}^{-}}-m_{\mathcal{B}_{c}^{+}})\bigr)-\bigl(m_{\mathcal{B}_{c}^{-}}^2+Q^2\bigr) (m_{\mathcal{B}_{b}^{*-}}+2 m_{\mathcal{B}_{c}^{+}}-2 m_{\mathcal{B}_{c}^{-}})\Bigg) - m_{\mathcal{B}_{b}^{*-}} \rho_{8}^{\mathrm{QCD-V}} \Bigg(2 m_{\mathcal{B}_{b}^{*+}}^2 (m_{\mathcal{B}_{b}^{*-}}-m_{\mathcal{B}_{c}^{+}}+m_{\mathcal{B}_{c}^{-}})\\
 & &+m_{\mathcal{B}_{b}^{*+}} m_{\mathcal{B}_{c}^{-}} (m_{\mathcal{B}_{b}^{*-}}+2 m_{\mathcal{B}_{c}^{+}}+m_{\mathcal{B}_{c}^{-}})
+3 m_{\mathcal{B}_{b}^{*+}} Q^2-\bigl(m_{\mathcal{B}_{c}^{-}}^2+Q^2\bigr) (m_{\mathcal{B}_{b}^{*-}}+2 m_{\mathcal{B}_{c}^{+}}-2 m_{\mathcal{B}_{c}^{-}})\Bigg) \\
 & &-2 m_{\mathcal{B}_{b}^{*+}} m_{\mathcal{B}_{b}^{*-}} \rho_{1}^{\mathrm{QCD-V}} (m_{\mathcal{B}_{b}^{*-}}+2 m_{\mathcal{B}_{c}^{+}}-2 m_{\mathcal{B}_{c}^{-}})
+2 m_{\mathcal{B}_{b}^{*+}} \rho_{3}^{\mathrm{QCD-V}} (m_{\mathcal{B}_{b}^{*-}}-m_{\mathcal{B}_{c}^{+}}+m_{\mathcal{B}_{c}^{-}})\\
 & &
-3 m_{\mathcal{B}_{b}^{*+}} \rho_{5}^{\mathrm{QCD-V}} (m_{\mathcal{B}_{b}^{*-}}-m_{\mathcal{B}_{c}^{+}}+m_{\mathcal{B}_{c}^{-}})
-3 m_{\mathcal{B}_{b}^{*+}} m_{\mathcal{B}_{b}^{*-}} \rho_{6}^{\mathrm{QCD-V}} (m_{\mathcal{B}_{b}^{*-}}+2 m_{\mathcal{B}_{c}^{+}}-2 m_{\mathcal{B}_{c}^{-}}) \Big\} \\
G_{3}^{++}(Q^2) & = & \frac{\text{exp}\big[\frac{m_{\mathcal{B}_{b}^{*+}}^{2}}{M_{1}^{2}}+\frac{m_{\mathcal{B}_{c}^{+}}^{2}}{M_{2}^{2}}\big]}
{\lambda_{\mathcal{B}_{b}^{*+}}\lambda_{\mathcal{B}_{c}^{+}} (m_{\mathcal{B}_{b}^{*+}}+m_{\mathcal{B}_{b}^{*-}}) (m_{\mathcal{B}_{c}^{+}}+m_{\mathcal{B}_{c}^{-}}) \bigl(m_{\mathcal{B}_{b}^{*+}} (m_{\mathcal{B}_{b}^{*-}}+m_{\mathcal{B}_{c}^{+}}-m_{\mathcal{B}_{c}^{-}})-(m_{\mathcal{B}_{c}^{+}}-m_{\mathcal{B}_{c}^{-}}) (m_{\mathcal{B}_{b}^{*-}}-2 m_{\mathcal{B}_{c}^{+}}+2 m_{\mathcal{B}_{c}^{-}})\bigr)}\\
 & &
\int^{s_{0}}_{s_{min}}ds\int^{u_{0}}_{u_{min}}du\,\text{exp}\big[-\frac{s}{M_1^2}-\frac{u}{M_2^2}\big]
\Big\{ m_{\mathcal{B}_{b}^{*+}} m_{\mathcal{B}_{c}^{+}} \rho_{11}^{\mathrm{QCD-V}} (-2 m_{\mathcal{B}_{b}^{*+}}^2 (m_{\mathcal{B}_{b}^{*-}}+m_{\mathcal{B}_{c}^{+}}-m_{\mathcal{B}_{c}^{-}})+m_{\mathcal{B}_{b}^{*+}} \bigl(3 m_{\mathcal{B}_{b}^{*-}}^2+m_{\mathcal{B}_{b}^{*-}} m_{\mathcal{B}_{c}^{-}}\\
 & &+m_{\mathcal{B}_{c}^{-}} (m_{\mathcal{B}_{c}^{+}}-m_{\mathcal{B}_{c}^{-}})\bigr)+\bigl(m_{\mathcal{B}_{c}^{-}}^2+Q^2\bigr) (m_{\mathcal{B}_{b}^{*-}}-2 m_{\mathcal{B}_{c}^{+}}+2 m_{\mathcal{B}_{c}^{-}})) + m_{\mathcal{B}_{b}^{*+}} m_{\mathcal{B}_{b}^{*-}} m_{\mathcal{B}_{c}^{+}} \rho_{9}^{\mathrm{QCD-V}} (2 m_{\mathcal{B}_{b}^{*+}}^2 (m_{\mathcal{B}_{b}^{*-}}+m_{\mathcal{B}_{c}^{+}}-m_{\mathcal{B}_{c}^{-}})\\
 & &+m_{\mathcal{B}_{b}^{*+}} m_{\mathcal{B}_{c}^{-}} (-m_{\mathcal{B}_{b}^{*-}}+2 m_{\mathcal{B}_{c}^{+}}+m_{\mathcal{B}_{c}^{-}})+3 m_{\mathcal{B}_{b}^{*+}} Q^2-\bigl(m_{\mathcal{B}_{c}^{-}}^2+Q^2\bigr) (m_{\mathcal{B}_{b}^{*-}}-2 m_{\mathcal{B}_{c}^{+}}+2 m_{\mathcal{B}_{c}^{-}})) \\
 & &+ 3 m_{\mathcal{B}_{b}^{*+}}^2 m_{\mathcal{B}_{b}^{*-}} m_{\mathcal{B}_{c}^{+}} \rho_{10}^{\mathrm{QCD-V}} (m_{\mathcal{B}_{b}^{*-}}-2 m_{\mathcal{B}_{c}^{+}}+2 m_{\mathcal{B}_{c}^{-}})
+ 3 m_{\mathcal{B}_{b}^{*+}}^2 m_{\mathcal{B}_{c}^{+}} \rho_{12}^{\mathrm{QCD-V}} (m_{\mathcal{B}_{b}^{*-}}+m_{\mathcal{B}_{c}^{+}}-m_{\mathcal{B}_{c}^{-}}) \\
 & &-2 m_{\mathcal{B}_{b}^{*+}} m_{\mathcal{B}_{c}^{+}} \rho_{1}^{\mathrm{QCD-V}} \bigl(3 m_{\mathcal{B}_{b}^{*+}} (m_{\mathcal{B}_{c}^{+}}-m_{\mathcal{B}_{c}^{-}})+m_{\mathcal{B}_{b}^{*-}} (m_{\mathcal{B}_{b}^{*-}}-2 m_{\mathcal{B}_{c}^{+}}+2 m_{\mathcal{B}_{c}^{-}})\bigr)
\\
 & &+2 m_{\mathcal{B}_{b}^{*+}} m_{\mathcal{B}_{c}^{+}} \rho_{3}^{\mathrm{QCD-V}} (m_{\mathcal{B}_{b}^{*-}}-2 m_{\mathcal{B}_{c}^{+}}+2 m_{\mathcal{B}_{c}^{-}}) \Big\} \\
G_{4}^{++}(Q^2) & = & \frac{\text{exp}\big[\frac{m_{\mathcal{B}_{b}^{*+}}^{2}}{M_{1}^{2}}+\frac{m_{\mathcal{B}_{c}^{+}}^{2}}{M_{2}^{2}}\big]}
{\lambda_{\mathcal{B}_{b}^{*+}}\lambda_{\mathcal{B}_{c}^{+}} (m_{\mathcal{B}_{b}^{*+}}+m_{\mathcal{B}_{b}^{*-}}) (m_{\mathcal{B}_{c}^{+}}+m_{\mathcal{B}_{c}^{-}}) \bigl(m_{\mathcal{B}_{b}^{*+}} (m_{\mathcal{B}_{b}^{*-}}+m_{\mathcal{B}_{c}^{+}}-m_{\mathcal{B}_{c}^{-}})-(m_{\mathcal{B}_{c}^{+}}-m_{\mathcal{B}_{c}^{-}}) (m_{\mathcal{B}_{b}^{*-}}-2 m_{\mathcal{B}_{c}^{+}}+2 m_{\mathcal{B}_{c}^{-}})\bigr)}\\
 & &
\int^{s_{0}}_{s_{min}}ds\int^{u_{0}}_{u_{min}}du\,\text{exp}\big[-\frac{s}{M_1^2}-\frac{u}{M_2^2}\big]
\Big\{ m_{\mathcal{B}_{c}^{+}}^2 \rho_{15}^{\mathrm{QCD-V}} (-2 m_{\mathcal{B}_{b}^{*+}}^2 (m_{\mathcal{B}_{b}^{*-}}+m_{\mathcal{B}_{c}^{+}}-m_{\mathcal{B}_{c}^{-}})+m_{\mathcal{B}_{b}^{*+}} \bigl(3 m_{\mathcal{B}_{b}^{*-}}^2+m_{\mathcal{B}_{b}^{*-}} m_{\mathcal{B}_{c}^{-}}\\
 & &+m_{\mathcal{B}_{c}^{-}} (m_{\mathcal{B}_{c}^{+}}-m_{\mathcal{B}_{c}^{-}})\bigr)+\bigl(m_{\mathcal{B}_{c}^{-}}^2+Q^2\bigr)  (m_{\mathcal{B}_{b}^{*-}}-2 m_{\mathcal{B}_{c}^{+}}+2 m_{\mathcal{B}_{c}^{-}}))+ m_{\mathcal{B}_{b}^{*-}} m_{\mathcal{B}_{c}^{+}}^2 \rho_{13}^{\mathrm{QCD-V}} (2 m_{\mathcal{B}_{b}^{*+}}^2 (m_{\mathcal{B}_{b}^{*-}}+m_{\mathcal{B}_{c}^{+}}-m_{\mathcal{B}_{c}^{-}})\\
 & &+m_{\mathcal{B}_{b}^{*+}} m_{\mathcal{B}_{c}^{-}} (-m_{\mathcal{B}_{b}^{*-}}+2 m_{\mathcal{B}_{c}^{+}}+m_{\mathcal{B}_{c}^{-}})+3 m_{\mathcal{B}_{b}^{*+}} Q^2-\bigl(m_{\mathcal{B}_{c}^{-}}^2+Q^2\bigr) (m_{\mathcal{B}_{b}^{*-}}-2 m_{\mathcal{B}_{c}^{+}}+2 m_{\mathcal{B}_{c}^{-}})) \\
 & &+ 3 m_{\mathcal{B}_{b}^{*+}} m_{\mathcal{B}_{b}^{*-}} m_{\mathcal{B}_{c}^{+}}^2 \rho_{14}^{\mathrm{QCD-V}} (m_{\mathcal{B}_{b}^{*-}}-2 m_{\mathcal{B}_{c}^{+}}+2 m_{\mathcal{B}_{c}^{-}})
+ 3 m_{\mathcal{B}_{b}^{*+}} m_{\mathcal{B}_{c}^{+}}^2 \rho_{16}^{\mathrm{QCD-V}} (m_{\mathcal{B}_{b}^{*-}}+m_{\mathcal{B}_{c}^{+}}-m_{\mathcal{B}_{c}^{-}}) \\
 & &+ 2 m_{\mathcal{B}_{b}^{*+}} m_{\mathcal{B}_{c}^{+}}^2 \rho_{7}^{\mathrm{QCD-V}} (m_{\mathcal{B}_{b}^{*-}}+m_{\mathcal{B}_{c}^{+}}-m_{\mathcal{B}_{c}^{-}})
- 2 m_{\mathcal{B}_{b}^{*+}} m_{\mathcal{B}_{b}^{*-}} m_{\mathcal{B}_{c}^{+}}^2 \rho_{8}^{\mathrm{QCD-V}} (m_{\mathcal{B}_{b}^{*-}}-2 m_{\mathcal{B}_{c}^{+}}+2 m_{\mathcal{B}_{c}^{-}}) \Big\}
\end{eqnarray*}
To eliminate the contributions from excited and continuum states, the threshold parameters $s_0$ and $u_0$ are introduced with their values conventionally taken as $s_0 = (m_{\mathcal B_{b}^*} + \Delta_1)^2$ and $u_0 = (m_{\mathcal B_{c}} + \Delta_2)^2$, where our previous work has shown that $\Delta_1 = \Delta_2 = 0.6 \sim 0.7$~GeV is reliable for heavy baryon masses~\cite{Wang:2010vn}. We therefore adopt $\Delta_1 = \Delta_2 = 0.6$~GeV to determine the central valueS of form factors, with an uncertainty of $0.6\pm 0.1$~GeV.

\begin{widetext}
\section{Helicity amplitudes}
\label{Sec:AppB}

Within the helicity formalism, the hadronic matrix elements are decomposed in terms of the helicity amplitudes $H_{\lambda_2,\lambda_W}$, where $\lambda_2$ and $\lambda_W$ denote the helicities of the final-state baryon and the virtual gauge boson, respectively. For the vector and axial-vector currents, the corresponding helicity amplitudes can be explicitly constructed as linear combinations of the form factors $F_i^V(q^2)$ and $G_i^A(q^2)$. We define the combinations $M_\pm = m_{\mathcal{B}_{b}^{*}} \pm m_{\mathcal{B}_c}$ and the kinematic factors $\alpha$ as follows,
\begin{align*}
\alpha_{\frac{1}{2}t}^{V} &= \alpha_{\frac{1}{2}0}^{A} = \sqrt{\frac{2 m_{\mathcal{B}_b^*} m_{\mathcal{B}_c} (\omega+1)}{q^2}}, \\
\alpha_{\frac{1}{2}0}^{V} &= \alpha_{\frac{1}{2}t}^{A} = \sqrt{\frac{2 m_{\mathcal{B}_b^*} m_{\mathcal{B}_c} (\omega-1)}{q^2}}, \\
\alpha_{\frac{1}{2}1}^{V} &= 2\sqrt{m_{\mathcal{B}_b^*} m_{\mathcal{B}_c} (\omega-1)}, \\
\alpha_{\frac{1}{2}1}^{A} &= 2\sqrt{m_{\mathcal{B}_b^*} m_{\mathcal{B}_c} (\omega+1)}.
\end{align*}

The helicity amplitudes can be expressed as following forms,
\begin{align}
H_{\frac{1}{2}t}^{V}
&= -\sqrt{\frac{2}{3}}\alpha^{V}_{\frac{1}{2}t}(\omega-1)
\Big[F_{1}^{V}(q^{2})m_{\mathcal{B}_c}
+F_{2}^{V}(q^{2})M_{+} -F_{3}^{V}(q^{2})\frac{m_{\mathcal{B}_b^{*}}}{m_{\mathcal{B}_c}}(m_{\mathcal{B}_b^{*}}-m_{\mathcal{B}_c}\omega)
-F_{4}^{V}(q^{2})\frac{q^{2}}{m_{\mathcal{B}_b^{*}}}\Big] \nonumber \\
H_{\frac{1}{2}0}^{V}
&= -\sqrt{\frac{2}{3}}\alpha^{V}_{\frac{1}{2}0}
\Big[F_{1}^{V}(q^{2})(m_{\mathcal{B}_b^{*}}-m_{\mathcal{B}_c}\omega)
+F_{2}^{V}(q^{2})(\omega+1)M_{-} -F_{3}^{V}(q^{2})(\omega^{2}-1)m_{\mathcal{B}_b^{*}}\Big] \nonumber \\
H_{\frac{1}{2}1}^{V}
&= \frac{1}{\sqrt{6}}\alpha^{V}_{\frac{1}{2}1}
\Big[F_{1}^{V}(q^{2})+2F_{2}^{V}(q^{2})(\omega+1)\Big] \nonumber \\
H_{\frac{1}{2}-1}^{V}
&= \frac{1}{\sqrt{2}}\alpha^{V}_{\frac{1}{2}1}F_{1}^{V}(q^{2}) \nonumber \\
H_{\frac{1}{2}t}^{A}
&= \sqrt{\frac{2}{3}}\alpha^{A}_{\frac{1}{2}t}(\omega+1)
\Big[G_{1}^{A}(q^{2})m_{\mathcal{B}_c}-G_{2}^{A}(q^{2})M_{-} -G_{3}^{A}(q^{2})\frac{m_{\mathcal{B}_b^{*}}}{m_{\mathcal{B}_c}}(m_{\mathcal{B}_b^{*}}-m_{\mathcal{B}_c}\omega)
-G_{4}^{A}(q^{2})\frac{q^{2}}{m_{\mathcal{B}_c}}\Big] \nonumber \\
H_{\frac{1}{2}0}^{A}
&= -\sqrt{\frac{2}{3}}\alpha^{A}_{\frac{1}{2}0}
\Big[-G_{1}^{A}(q^{2})(m_{\mathcal{B}_b^{*}}-m_{\mathcal{B}_c}\omega)
+G_{2}^{A}(q^{2})(\omega-1)M_{+} +G_{3}^{A}(q^{2})(\omega^{2}-1)m_{\mathcal{B}_b^{*}}\Big] \nonumber \\
H_{\frac{1}{2}1}^{A}
&= \frac{1}{\sqrt{6}}\alpha^{A}_{\frac{1}{2}1}
\Big[-G_{1}^{A}(q^{2})+2G_{2}^{A}(q^{2})(\omega-1)\Big] \nonumber \\
H_{\frac{1}{2}-1}^{A}
&= -\frac{1}{\sqrt{2}}\alpha^{A}_{\frac{1}{2}1}G_{1}^{A}(q^{2}) \label{eq:B1}
\end{align}

The amplitudes for negative helicity satisfy the relations,
\begin{equation*}
H^{V}_{-\lambda_{2},-\lambda_{W}} = -H^{V}_{\lambda_{2},\lambda_{W}} \qquad
H^{A}_{-\lambda_{2},-\lambda_{W}} = H^{A}_{\lambda_{2},\lambda_{W}}
\end{equation*}

The squared amplitude $|M|^2$ is given by,
\begin{align}\notag
\left|M\right|^2 =
&\frac{2}{3}\,(q^2 - m_l^2)\Bigg\{
\frac{3}{8}(1-\cos\theta)^2 \left( |H_{\frac{1}{2}1}|^2 + |H_{-\frac{1}{2}1}|^2 \right)
+ \frac{3}{8}(1+\cos\theta)^2 \left( |H_{-\frac{1}{2}-1}|^2 + |H_{\frac{1}{2}-1}|^2 \right)   \\ \notag
&+ \frac{3}{4}\sin^2\theta \left( |H_{\frac{1}{2}0}|^2 + |H_{-\frac{1}{2}0}|^2 \right)
+ \frac{m_l^2}{2q^2} \Bigg[
\frac{3}{2}\left( |H_{\frac{1}{2}t}|^2 + |H_{-\frac{1}{2}t}|^2 \right)
+ \frac{3}{4}\sin^2\theta \left( |H_{\frac{1}{2}1}|^2 + |H_{-\frac{1}{2}-1}|^2 + |H_{-\frac{1}{2}1}|^2 + |H_{\frac{1}{2}-1}|^2 \right) \\
&+ \frac{3}{2}\cos^2\theta \left( |H_{\frac{1}{2}0}|^2 + |H_{-\frac{1}{2}0}|^2 \right)
- 3\cos\theta \left( H_{\frac{1}{2}t}H_{\frac{1}{2}0} + H_{-\frac{1}{2}t}H_{-\frac{1}{2}0} \right)
\Bigg] \Bigg\}
\label{eq:B2}
\end{align}

\end{widetext}

\begin{thebibliography}{}
\bibitem{ParticleDataGroup:2026mpi}
F.~Takahashi \textit{et al.} [Particle Data Group],
\href{https://doi.org/10.1142/s0217751x26300115}{Int. J. Mod. Phys. A \textbf{41}, 2630011 (2026)}.
\bibitem{UA1:1991vse}
C.~Albajar \textit{et al.} [UA1],
\href{https://doi.org/10.1016/0370-2693(91)90311-D}{Phys. Lett. B \textbf{273}, 540 (1991)}.
\bibitem{D0:2007gjs}
V.~M.~Abazov \textit{et al.} [D0],
\href{https://doi.org/10.1103/PhysRevLett.99.052001}{Phys. Rev. Lett. \textbf{99}, 052001 (2007)}.
\bibitem{D0:2008sbw}
V.~M.~Abazov \textit{et al.} [D0],
\href{https://doi.org/10.1103/PhysRevLett.101.232002}{Phys. Rev. Lett. \textbf{101}, 232002 (2008)}.
\bibitem{CMS:2012frl}
S.~Chatrchyan \textit{et al.} [CMS],
\href{https://doi.org/10.1103/PhysRevLett.108.252002}{Phys. Rev. Lett. \textbf{108}, 252002 (2012)}.
\bibitem{CDF:2007oeq}
T.~Aaltonen \textit{et al.} [CDF],
\href{https://doi.org/10.1103/PhysRevLett.99.202001}{Phys. Rev. Lett. \textbf{99}, 202001 (2007)}.
\bibitem{Ebert:2011kk}
D.~Ebert, R.~N.~Faustov and V.~O.~Galkin,
\href{https://doi.org/10.1103/PhysRevD.84.014025}{Phys. Rev. D \textbf{84}, 014025 (2011)}.
\bibitem{Capstick:1986ter}
S.~Capstick and N.~Isgur,
\href{https://doi.org/10.1103/physrevd.34.2809}{Phys. Rev. D \textbf{34}, 2809 (1986)}.
\bibitem{Roberts:2007ni}
W.~Roberts and M.~Pervin,
\href{https://doi.org/10.1142/S0217751X08041219}{Int. J. Mod. Phys. A \textbf{23}, 2817 (2008)}.
\bibitem{Garcilazo:2007eh}
H.~Garcilazo, J.~Vijande and A.~Valcarce,
\href{https://doi.org/10.1088/0954-3899/34/5/014}{J. Phys. G \textbf{34}, 961 (2007)}.
\bibitem{Wang:2010fq}
Z.~G.~Wang,
\href{https://doi.org/10.1140/epjc/s10052-010-1365-8}{Eur. Phys. J. C \textbf{68}, 479 (2010)}.
\bibitem{Chen:2019ywy}
B.~Chen, S.~Q.~Luo, X.~Liu and T.~Matsuki,
\href{https://doi.org/10.1103/PhysRevD.100.094032}{Phys. Rev. D \textbf{100},  094032 (2019)}.
\bibitem{Wang:2020mxk}
Z.~G.~Wang and H.~J.~Wang,
\href{https://doi.org/10.1088/1674-1137/abc1d3}{Chin. Phys. C \textbf{45}, 013109 (2021)}.
\bibitem{Wang:2009cr}
Z.~G.~Wang,
\href{https://doi.org/10.1016/j.physletb.2010.01.039}{Phys. Lett. B \textbf{685}, 59 (2010)}.

























\bibitem{Efimov:1991qj}
G.~V.~Efimov, M.~A.~Ivanov, N.~B.~Kulimanova and V.~E.~Lyubovitskij,
\href{https://doi.org/10.1007/BF01566666}{Z. Phys. C \textbf{54}, 349 (1992)}.
\bibitem{Zhang:2026ugd}
S.~W.~Zhang, X.~Luo, H.~X.~Chen and H.~M.~Yang,
\href{https://arxiv.org/abs/2608.18920}{arXiv:2608.18920 [hep-ph]}.
\bibitem{Niu:2026fzi}
P.~Y.~Niu, Q.~Wang and Q.~Zhao,
\href{https://doi.org/10.1088/0256-307X/43/7/070202}{Chin. Phys. Lett. \textbf{43}, 070202 (2026)}.
\bibitem{Luo:2025hjx}
X.~Luo, H.~M.~Yang and H.~X.~Chen,
\href{https://doi.org/10.22323/1.500.0036}{PoS \textbf{HADRON2025}, 036 (2026)}.
\bibitem{Wang:2025amv}
Y.~J.~Wang, X.~Luo, H.~X.~Chen, E.~L.~Cui, W.~H.~Tan and Z.~Y.~Zhou,
\href{https://doi.org/10.22323/1.500.0037}{PoS \textbf{HADRON2025}, 037 (2026)}.

\bibitem{Wang:2024rai}
Y.~J.~Wang, X.~Luo, H.~X.~Chen, E.~L.~Cui, W.~H.~Tan and Z.~Y.~Zhou,
\href{https://doi.org/10.1103/PhysRevD.111.076003}{Phys. Rev. D \textbf{111}, 076003 (2025)}.
\bibitem{Shu:2024jdn}
Y.~L.~Shu, Q.~F.~Song and Q.~F.~L{\"u},
\href{https://doi.org/10.1088/1572-9494/adf838}{Commun. Theor. Phys. \textbf{78}, 015301 (2026)}.
\bibitem{Negash:2026orr}
H.~Negash, G.~Tekle, M.~Gebremichael and G.~Seyoum,
\href{https://doi.org/10.1142/s0218301326500242}{Int. J. Mod. Phys. E \textbf{35}, 2650024 (2026)}.
\bibitem{Patel:2026skc}
R.~V.~Patel, M.~Shah and B.~Pandya,
\href{https://doi.org/10.1140/epjc/s10052-026-15778-x}{Eur. Phys. J. C \textbf{86}, 608 (2026)}.
\bibitem{Rivero-Acosta:2025drn}
A.~Rivero-Acosta, H.~Garc{\'\i}a-Tecocoatzi, A.~Ramirez-Morales, E.~Santopinto and C.~A.~Vaquera-Araujo,
\href{https://doi.org/10.1103/ng36-kn1p}{Phys. Rev. D \textbf{112}, 072014 (2025)}.
\bibitem{Aliev:2009jt}
T.~M.~Aliev, K.~Azizi and A.~Ozpineci,
\href{https://doi.org/10.1103/PhysRevD.79.056005}{Phys. Rev. D \textbf{79}, 056005 (2009)}.
\bibitem{Yao:2018jmc}
Y.~X.~Yao, K.~L.~Wang and X.~H.~Zhong,
\href{https://doi.org/10.1103/PhysRevD.98.076015}{Phys. Rev. D \textbf{98}, 076015 (2018)}.

\bibitem{Wang:2017kfr}
K.~L.~Wang, Y.~X.~Yao, X.~H.~Zhong and Q.~Zhao,
\href{https://doi.org/10.1103/PhysRevD.96.116016}{Phys. Rev. D \textbf{96}, 116016 (2017)}.
\bibitem{Meinel:2021rbm}
S.~Meinel and G.~Rendon,
\href{https://doi.org/10.1103/PhysRevD.103.094516}{Phys. Rev. D \textbf{103}, 094516 (2021)}.


\bibitem{Detmold:2015aaa}
W.~Detmold, C.~Lehner and S.~Meinel,
\href{https://doi.org/10.1103/PhysRevD.92.034503}{Phys. Rev. D \textbf{92}, 034503 (2015)}.
\bibitem{Lu:2026qkk}
J.~Lu, G.~L.~Yu, D.~Y.~Chen, Z.~G.~Wang and B.~Wu,
\href{https://doi.org/10.1140/epjc/s10052-026-15871-1}{Eur. Phys. J. C \textbf{86}, 685 (2026)}.
\bibitem{Ke:2019smy}
H.~W.~Ke, N.~Hao and X.~Q.~Li,
\href{https://doi.org/10.1140/epjc/s10052-019-7048-1}{Eur. Phys. J. C \textbf{79}, 540 (2019)}.
\bibitem{Wang:2022ias}
W.~Wang and Z.~P.~Xing,
\href{https://doi.org/10.1016/j.physletb.2022.137402}{Phys. Lett. B \textbf{834}, 137402 (2022)}.
\bibitem{Aliev:2010uy}
T.~M.~Aliev, K.~Azizi and M.~Savci,
\href{https://doi.org/10.1103/PhysRevD.81.056006}{Phys. Rev. D \textbf{81}, 056006 (2010)}.
\bibitem{Azizi:2011mw}
K.~Azizi, Y.~Sarac and H.~Sundu,
\href{https://doi.org/10.1140/epja/i2012-12002-1}{Eur. Phys. J. A \textbf{48}, 2 (2012)}.
\bibitem{Ebert:2006rp}
D.~Ebert, R.~N.~Faustov and V.~O.~Galkin,
\href{https://doi.org/10.1103/PhysRevD.73.094002}{Phys. Rev. D \textbf{73}, 094002 (2006)}.
\bibitem{Cheng:1995fe}
H.~Y.~Cheng and B.~Tseng,
\href{https://doi.org/10.1103/PhysRevD.53.1457}{Phys. Rev. D \textbf{53}, 1457 (1996)}.
\bibitem{Wang:2010it}
Z.~G.~Wang,
\href{https://doi.org/10.1140/epja/i2011-11081-8}{Eur. Phys. J. A \textbf{47}, 81 (2011)}.
\bibitem{Neishabouri:2025abl}
Z.~Neishabouri and K.~Azizi,
\href{https://doi.org/10.1103/p1s1-dhqd}{Phys. Rev. D \textbf{112}, 054009 (2025)}.
\bibitem{Neishabouri:2024gbc}
Z.~Neishabouri, K.~Azizi and H.~R.~Moshfegh,
\href{https://doi.org/10.1103/PhysRevD.110.014010}{Phys. Rev. D \textbf{110}, 014010 (2024)}.
\bibitem{Luo:2025sns}
X.~Luo, S.~W.~Zhang, H.~X.~Chen, A.~Hosaka, N.~Su and H.~M.~Yang,
\href{https://doi.org/10.1007/s43673-026-00188-8}{AAPPS Bull. \textbf{36}, 9 (2026)}.
\bibitem{Lu:2025gol}
J.~Lu, G.~L.~Yu, D.~Y.~Chen, Z.~G.~Wang and B.~Wu,
\href{https://doi.org/10.1140/epjc/s10052-025-15110-z}{Eur. Phys. J. C \textbf{85}, 1382 (2025)}.
\bibitem{Yu:2026tbk}
G.~L.~Yu, Z.~G.~Wang, J.~Lu, B.~Wu, P.~Yang and Z.~Zhou,
\href{https://doi.org/10.1103/ghm9-2lmd}{Phys. Rev. D \textbf{113}, 7 (2026)}.
\bibitem{Wang:2026eci}
Z.~G.~Wang and Y.~Liu,
\href{https://arxiv.org/abs/2607.17492}{arXiv:2607.17492 [hep-ph]}.
\bibitem{Wang:2026dqi}
Z.~G.~Wang and Y.~Liu,
\href{https://doi.org/10.15302/frontphys.2027.016201}{Front. Phys. (Beijing) \textbf{22}, 016201 (2027)}.
\bibitem{Zhou:2025yjb}
Z.~Zhou, G.~L.~Yu, Z.~G.~Wang, J.~Lu and B.~Wu,
\href{https://doi.org/10.1140/epja/s10050-026-01805-8}{Eur. Phys. J. A \textbf{62}, 39 (2026)}.
\bibitem{Khajouei:2025tqw}
L.~Khajouei and K.~Azizi,
\href{https://doi.org/10.1016/j.nuclphysb.2026.117560}{Nucl. Phys. B \textbf{1029}, 117560 (2026)}.
\bibitem{Khajouei:2024frw}
L.~Khajouei and K.~Azizi,
\href{https://doi.org/10.1103/PhysRevD.111.074018}{Phys. Rev. D \textbf{111}, 074018 (2025)}.
\bibitem{Khodjamirian:2020btr}
A.~Khodjamirian,
\href{https://doi.org/10.1201/9781315142005}{CRC Press, (2020)}.
\bibitem{Colangelo:2000dp}
P.~Colangelo and A.~Khodjamirian
\href{https://arxiv.org/abs/hep-ph/0010175}{arXiv:hep-ph/0010175 [hep-ph]}.
\bibitem{Shifman:1978bx}
M.~A.~Shifman, A.~I.~Vainshtein and V.~I.~Zakharov,
\href{https://doi.org/10.1016/0550-3213(79)90022-1}{Nucl. Phys. B \textbf{147}, 385 (1979)}.
\bibitem{Shifman:1978by}
M.~A.~Shifman, A.~I.~Vainshtein and V.~I.~Zakharov,
\href{https://doi.org/10.1016/0550-3213(79)90023-3}{Nucl. Phys. B \textbf{147}, 448 (1979)}.
\bibitem{Cutkosky:1960sp}
R.~E.~Cutkosky,
\href{https://doi.org/10.1063/1.1703676}{J. Math. Phys. \textbf{1}, 429 (1960)}.
\bibitem{Bracco:2011pg}
M.~E.~Bracco, M.~Chiapparini, F.~S.~Navarra and M.~Nielsen,
\href{https://doi.org/10.1016/j.ppnp.2012.03.002}{Prog. Part. Nucl. Phys. \textbf{67}, 1019 (2012)}.
\bibitem{ParticleDataGroup:2024cfk}
S.~Navas \textit{et al.} [Particle Data Group],
\href{https://doi.org/10.1103/PhysRevD.110.030001}{Phys. Rev. D \textbf{110}, 030001 (2024)}.
\bibitem{Yu:2022ymb}
G.~L.~Yu, Z.~Y.~Li, Z.~G.~Wang, J.~Lu and M.~Yan,
\href{https://doi.org/10.1016/j.nuclphysb.2023.116183}{Nucl. Phys. B \textbf{990}, 116183 (2023)}.
\bibitem{Li:2024zze}
Z.~Y.~Li, G.~L.~Yu, Z.~G.~Wang and J.~Z.~Gu,
\href{https://doi.org/10.1140/epjc/s10052-024-13706-5}{Eur. Phys. J. C \textbf{84}, 1310 (2024)}.

\bibitem{Wang:2010vn}
Z.~G.~Wang,
\href{https://doi.org/10.1140/epjc/s10052-010-1357-8}{Eur. Phys. J. C \textbf{68}, 459 (2010)}.
\bibitem{Lee:2022jjn}
H.~J.~Lee,
\href{https://doi.org/10.3938/NPSM.72.887}{New Phys. Sae Mulli \textbf{72}, 887 (2022)}.
\bibitem{Reinders:1984sr}
L.~J.~Reinders, H.~Rubinstein and S.~Yazaki,
\href{https://doi.org/10.1016/0370-1573(85)90065-1}{Phys. Rept. \textbf{127}, 1 (1985)}.
\bibitem{Narison:2010cg}
S.~Narison,
\href{https://doi.org/10.1016/j.physletb.2011.09.116}{Phys. Lett. B \textbf{693}, 559 (2010)}.
\bibitem{Narison:2011xe}
S.~Narison,
\href{https://doi.org/10.1016/j.physletb.2011.11.058}{Phys. Lett. B \textbf{706}, 412 (2012)}.
\bibitem{Narison:2011rn}
S.~Narison,
\href{https://doi.org/10.1016/j.physletb.2011.12.047}{Phys. Lett. B \textbf{707}, 259 (2012)}.
\bibitem{Li:2023wgq}
X.~J.~Li, Y.~S.~Li, F.~L.~Wang and X.~Liu,
\href{https://doi.org/10.1140/epjc/s10052-023-12237-9}{Eur. Phys. J. C \textbf{83}, 1080 (2023)}.
\bibitem{Kadeer:2005aq}
A.~Kadeer, J.~G.~Korner and U.~Moosbrugger,
\href{https://doi.org/10.1140/epjc/s10052-008-0801-5}{Eur. Phys. J. C \textbf{59}, 27 (2009)}.
\bibitem{Wang:2010vn}
Z.~G.~Wang,
\href{https://doi.org/10.1140/epjc/s10052-010-1357-8}{Eur. Phys. J. C \textbf{68}, 459 (2010)}.


















\end{thebibliography}
\end{document}